\documentclass[twocolumn]{aastex63}

\newcommand\nustar{{\it NuSTAR}}
\newcommand\suzaku{{\it Suzaku}}
\newcommand\xmm{{\it XMM-Newton}}
\newcommand\chandra{{\it Chandra}}
\newcommand\planck{{\it Planck}}
\newcommand\eros{{\it eROSITA}}
\usepackage{tikz}
\usetikzlibrary{shapes,arrows}
\tikzstyle{block} = [rectangle, draw, fill=red!20, 
    text width=25em, text centered, rounded corners, minimum height=3em]
\tikzstyle{line} = [draw, -latex']
\received{2021 December 10}
\revised{2022 February 28}
\accepted{2022 March 27}
\submitjournal{ApJ}

\shorttitle{The \nustar, \xmm, and \suzaku\ view of A3395}
\shortauthors{T\"{u}mer et al.}

\begin{document}

\title{The NuSTAR, XMM-Newton, and Suzaku view of A3395 at the intercluster filament interface}

\correspondingauthor{Ay\c{s}eg\"{u}l T\"{u}mer}
\email{aysegultumer@gmail.com}

\author[0000-0002-3132-8776]{Ay\c{s}eg\"{u}l T\"{u}mer}
\affiliation{Department of Physics \& Astronomy,
University of Utah,
115 South 1400 East, Salt Lake City, UT 84112, USA}

\author[0000-0001-9110-2245]{Daniel R. Wik}
\affiliation{Department of Physics \& Astronomy,
University of Utah,
115 South 1400 East, Salt Lake City, UT 84112, USA}

\author[0000-0003-2754-9258]{Massimo Gaspari}
\affiliation{INAF, Osservatorio di Astrofisica e Scienza dello Spazio, via Pietro Gobetti 93/3, 40129 Bologna, Italy}
\affiliation{Department of Astrophysical Sciences, Princeton University, 4 Ivy Lane, Princeton, NJ 08544-1001, USA}

\author[0000-0003-1949-7005]{Hiroki Akamatsu}
\affiliation{SRON Netherlands Institute for Space Research, Niels Bohrweg 4, 2333 CA Leiden, The Netherlands}

\author[0000-0001-5839-8590]{Niels J. Westergaard}
\affiliation{DTU Space, Technical University of Denmark,
Elektrovej Building 327, DK-2800
Kgs Lyngby, Denmark}

\author[0000-0002-6562-8654]{Francesco Tombesi}
\affiliation{Department of Physics, University of Rome ``Tor Vergata", Via della Ricerca Scientifica 1, 00133 Rome, Italy}
\affiliation{Department of Astronomy, University of Maryland, College Park, MD 20742, USA}
\affiliation{NASA/Goddard Space Flight Center, Code 662, Greenbelt, MD 20771, USA}
\affiliation{INAF Osservatorio Astronomico di Roma, Via Frascati 33, 00078 Monteporzio Catone, Italy}

\author[0000-0003-0639-7048]{E. Nihal Ercan}
\affiliation{Department of Physics, Bo\u{g}azi\c{c}i University, Bebek, 34342 Istanbul, Turkey}

\begin{abstract}

Galaxy clusters are the largest virialized objects in the universe. Their merger dynamics and their interactions with the cosmic filaments that connect them are important for our understanding of the formation of large-scale structure. In addition, cosmic filaments are thought to possess the missing baryons in the universe. Studying the interaction of galaxy clusters and filaments therefore has the potential to unveil the the origin of the baryons and the physical processes that occur during merger stages of galaxy clusters.
In this paper, we study the connection between A3395 and the intercluster filament with \nustar, \xmm, and \suzaku~data. Since the \nustar\ observation is moderately contaminated by scattered light, we present a novel technique developed for disentangling this background from the emission from the intracluster medium. We find that the interface of the cluster and the intercluster filament connecting A3395 and A3391 does not show any signs of heated plasma, as was previously thought. This interface has low temperature, high density, and low entropy, thus we suggest that the gas is cooling, being enhanced by the turbulent or tidal `weather' driven during the early stage of the merger. Furthermore, our temperature results from the \nustar\ data are in agreement with those from \xmm, and from joint \nustar\ and \xmm\ analysis for a region with $\sim$25\% scattered light contamination within 1$\sigma$. We show that the temperature constraint of the intracluster medium is valid even when the data are contaminated up to $\sim$25\% for $\sim$5 keV cluster emission.

\end{abstract}



\section{Introduction} \label{sec:intro}
Clusters of galaxies are the largest gravitationally bound structures in the universe. Elements produced inside a galaxy cluster can rarely escape its deep gravitational potential well; therefore these constituents make clusters great probes for understanding the evolutionary history of large-scale structures. The intracluster medium (ICM) is an optically thin hot plasma ($\sim$10$^{7}$-10$^{8}$ K) that fills the volume between cluster galaxies.
It accounts for $\sim$12\% of the total matter inside galaxy clusters and its emission prevails in the X-ray band of the electromagnetic spectrum \citep[see, e.g.][]{sarazin86}. X-ray emitting processes in the ICM are mainly in the form of thermal bremsstrahlung and line emission. The plasma is close to hydrostatic equilibrium in relaxed clusters with a smooth, centrally peaked X-ray surface brightness distribution. However, a significant number of clusters show multipeaked brightness distributions, pointing to multiple substructures that indicate ongoing merger activity \citep{nakamura95}. 

\begin{figure*}
\centering
\includegraphics[width=150mm]{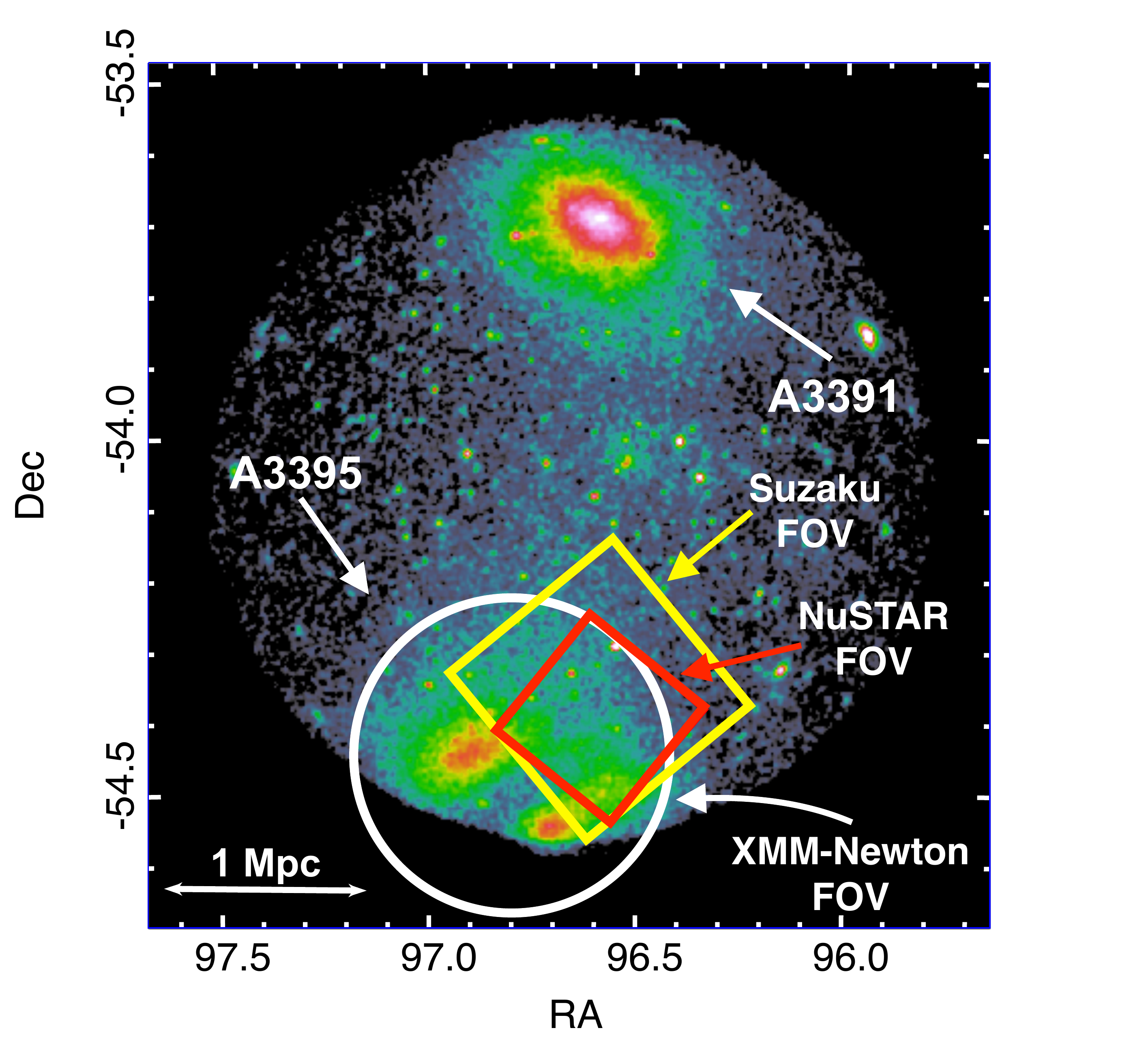}
\caption{\eros~image of Abell 3395 and Abell 3391 system. \nustar, \xmm, and \suzaku~FOV are indicated with a red box, white circle, and yellow box on the image, respectively. \label{fig:erositaphoton}}
\end{figure*}

According to scenarios for the formation of large-scale structure, clusters of galaxies are hierarchically formed by the merger of smaller-scale structures. The collisions of these substructures happen at very high velocities ($\sim$2000 km s$^{-1}$) and the energy released is of the order of 10$^{65}$~erg, making them the second most energetic events in the universe, following the Big Bang. This energy mostly is dissipated into the ICM through the creation of turbulence, magnetic fields, and relativistic particles (e.g., \citealt{markevitch99,gaspari13}), and it is ultimately thermalized.

Major mergers between clusters of comparable masses produce giant shock waves that can be directly detected in the density and temperature structure of the gas, and both major and minor mergers can disturb the cool gas at cluster centers, causing it to ``slosh" and produce cold fronts
\citep[see, e.g.,][for a review]{markevitch07}. 
Shock heating is indirectly evidenced by high entropy regions inside clusters \citep[see, e.g.,][]{henry95}, and shock fronts create sharp surface brightness discontinuities along with temperature jumps characterized by a Mach number $\la$3 
\citep{markevitch07}.
The boundary between a cluster and the intergalactic medium is defined by an accretion shock, where cooler gas meets the hotter gas of the ICM, which lies beyond the virial radius.
In contrast, filaments can penetrate more deeply into clusters, potentially creating shocks in locations where the gas is more easily observable in X-rays
\citep[e.g.,][]{zinger16}.

The existence of substructures in clusters of galaxies, since dynamical evolution is prone to wipe them out, points to dynamically young systems, i.e., clusters at a pre-merger stage \citep{flin03}. Thermodynamical properties of the X-ray emitting ICM are sensitive probes of these dynamical activities; moreover, in conjunction with the coevolving central supermassive black holes, such processes are key to shaping ICM X-ray scaling relations (\citealt{gaspari19,lovisari21}).

Clusters of galaxies are not completely isolated structures but are connected to the web-like structure of the matter distribution in the universe. This cosmic web consists of sheets, filaments, knots, and voids, where galaxies form and evolve \citep{bond96}. About $\sim$50\% of the baryonic matter lies in filaments, which only constitute $\sim$6\% of the volume \citep{cautun14}, making filaments relatively high-density structures, and galaxy clusters are found where they intersect. They are the end product of gravitational collapse of matter, where the baryonic gas follows the dark matter distribution \citep[see, e.g.,][]{hahn07,codis12,laigle15,kraljic18} that grew from small overdensities in the early universe. Clusters of galaxies sit at the highest density regions of this web, making cluster outskirts an inherently interesting region for the understanding of the formation, evolution, and cosmology of large-scale structures \citep{kuchner20}.

At a redshift $z = 0.0498$, Abell 3395 (hereafter A3395) extends out to a virial radius r$_{180}$=34$\arcmin$.6 \citep{markevitch98} and is classified as a merging clusters of galaxies, with its subclusters estimated to be near their first core passage \citep{lakhchaura11}. These subclusters are found to be close in redshift, indicating that the merger is taking place in the plane of the sky. However, the structure of the cluster seems to be even more complex, with two more relatively strong surface brightness peaks between the subclusters and an additional surface brightness excess to the west \citep{lakhchaura11}. Two scenarios are proposed for the origin of the surface brightness peaks in these regions by \citet{lakhchaura11}. The first scenario considers the western (W) subclump being a separate system that is in the first stage of a merger and falling into the common gravitational potential of the northeastern (NE) and southwestern (SW) subclusters. This scenario unfortunately does not explain the existence of the bridge (B) structure between the NE and SW subclusters. Another possibility is that we are witnessing the aftermath of a merger between the subclusters that has already occurred, with the bridge and W subclump resulting from ram pressure stripping of gas from the SW subcluster, possibly in two different phases. Moreover, the lack of prominent cool cores at the centers of the NE and SW subclusters are particularly interesting if these structures have not yet gone through a merger, since mergers are assumed to be the main processes that disrupt cool cores, given that feedback by central active galactic nuclei (AGNs) is expected to be self-regulated and gentle (\citealt{gaspari20}). In addition, diffuse radio structures lying in the W subclump have recently been discovered \citep{erosita21}. 

In addition, the cluster is part of a larger network of clusters and groups, where an emission bridge connecting A3395 to Abell 3391 (hereafter A3391) has also discovered along with a group of galaxies lying in between A3395 and A3391 \citep{tittley01,planck13,sugawara17,alvarez18}. In this work, we refer to this emission bridge as ``the intercluster filament". The intercluster filament has also recently been confirmed and  studied in detail with {\it eROSITA} \citep{erosita21}. Furthermore, this network has been mapped by \chandra\ and through the thermal Sunyaev-Zel'dovich (SZ) effect using the \planck~High Frequency Instrument by \citet{bourdin20}, which clearly shows the intercluster filament connected to the ICM of both clusters. These clusters are thought to have already begun interacting via this filament and might be at a pre-merger stage \citep{sugawara17}.

The excess emission is thought to originate from the interaction of A3395 and A3391, which resulted in tidally stripped cluster gas because of the lack of detection of warm ({\it kT} $<$ 1 keV) gas \citep{sugawara17,alvarez18}. This intercluster filament emission, which extends up to 3 Mpc, could not be fully explained by the galaxy group emission alone, but hints at the existence of warm and primordial filamentary gas.

\citet{alvarez18} suggest that the intercluster filament region is filled with tidally removed ICM from A3395 and A3391, evidenced by the filament temperature and entropy, which suggests a pre-merger stage. They also state that, although the global temperature ($kT = 4.45^{+0.89}_{-0.55}$~keV) and entropy profiles are higher than what is expected for the warm-hot intergalactic medium (WHIM) at the redshift \citep{valageas03}, while the density ($n_{e} = 1.08^{+0.06}_{-0.05} \times  10^{-4}$~cm$^{-3}$) of the filament is in agreement with the WHIM as are the cluster outskirts.

Also, a hot spot ($kT \simeq 9$~keV) was detected with \xmm\ in the NW region of A3395 by \citet{lakhchaura11}, and they suggest that this region of the cluster is part of the intercluster filament.

A399-A401 system has the other \planck~-detected filament \citep{planck13}. \citet{bonjean18} favor the scenario in which the emission bridge between A399-A401 is associated with a cosmic filament, where the gas is collapsing and being heated by the future merger of the these clusters. \citet{akamatsu17} and \citet{hincks22} discuss the effects of projection on the temperature and density estimations. Other similar galaxy cluster pre-mergers with an intercluster filament are seen in the A21-PSZ2 G114.90-34.35 pair \citep{bonjean18}, Shapley Supercluster (connecting A3562, A3558 and A3556 \citep{kull99}, and A222-A223 pair \citep{werner08}. The Shapley supercluster was observed with {\it ROSAT}. The authors debate that the origin of the excess emission is either overlapping gas distributions of A3558 and A3556 or an intrasupercluster emission \citep{kull99}. \citet{werner08} claim that the emission bridge connecting A222-A223 is the hottest and densest phase of the WHIM, due to the temperature and average density of the observed gas.

\begin{deluxetable*}{lccccc}
\tabletypesize{\scriptsize}
\tablewidth{0pt} 
\tablecaption{Observation log. \label{tab:obslog}}
\tablehead{
\colhead{Telescope} & \colhead{Observation ID} & \colhead{Start Date} & \colhead{PI} &\colhead{Equatorial coordinates}  & \colhead{Total Effective} \\
[-0.95em]
\colhead{} & \colhead{} & \colhead{(YYYY-mm-dd)} & \colhead{} & \colhead{(J2000)} & \colhead{Exposure time (ks)\tablenotemark{a}}
}
\startdata
\nustar & 70601003002 & 2020-09-09 & A. T\"{u}mer & 06:26:22, -54:23:47 & 250.4 (97.4\%) \\
\xmm & 0400010301 & 2007-01-24  & M. Henriksen & 06:27:11, -54:27:59 & \phn78.9 (90.4\%)\\
\suzaku & 807031010 & 2013-02-06  & N. Tanaka & 06:26:26, -54:20:22 & 100.5 (82.5\%)\\
\enddata
\tablenotetext{a}{Combined instrument exposures, i.e., FPMA and FPMB for \nustar; MOS1, MOS2 and PN for \xmm; and XIS1, XIS2 and XIS3 for \suzaku. Percentages correspond to the accepted data after filtering, with respect to the total raw exposure time.}
\end{deluxetable*}

Our main objective in this work is to understand the ICM and intercluster filament interaction at the junction of the cluster outskirts and the intercluster filament, which is different from the aforementioned studies that focus on the emission bridges. To achieve this, high-precision temperature measurement of the this region is required to assess a possible heating at the site. The peak effective area of \nustar\ \citep[Nuclear Spectroscopic Telescope Array;][]{harrison13} at E $\sim{10}$ keV and its 3 - 79 keV operating energy band enable the detection of recently shocked gas. Given that the hot spot seems to be at a temperature $\sim$9 keV \citep{lakhchaura11}, \nustar\ is the only observatory that is capable of making the measurement with highest precision required.

In this paper, we present results from the analysis of \nustar\, \xmm\ and \suzaku\ data to study the interaction of A3395 outskirts with the intercluster filament that connects A3395 and A3391.

The paper is organized as follows: observations, the data reduction process, and the background assessment of the \nustar, \xmm, and \suzaku\ data are presented in Section~\ref{sec:reduction}. In Section~\ref{sec:analysis}, methods used for the analysis of the cluster are described, including the description of our technique to treat the scattered light, and we present our results. We then discuss our findings in Section~\ref{sec:discussion}, and conclude our work in Section~\ref{sec:conclusion}.

Throughout this paper, we assume the $\Lambda$CDM cosmology with {\it H$_{0}$} = 70 km s$^{-1}$ Mpc$^{-1}$, $\Omega$$_{M}$ = 0.3, $\Omega$$_{\Lambda}$ = 0.7. According to these assumptions, a projected intracluster distance of 100 kpc corresponds to an angular separation of $\sim$103$\arcsec$ at the redshift of A3395, which makes the conversion from angular distance to physical distance straightforward for studying the images. 
The redshift value is fixed at $z = 0.0498$ (SIMBAD\footnote{\url{http://simbad.u-strasbg.fr/simbad/}} astronomical database), and we adopt a neutral hydrogen column density of $N_{H} = 6.30 \times 10^{20}$~cm$^{-2}$ based on Leiden/Argentine/Bonn Galactic HI survey \citep{kalberla05}. We used the abundance table of Anders \& Grevesse \citep{anders89} for \xmm~since the \xmm~ analyses in the literature were made using older tables. For \suzaku~analysis we used protosolar abundances \citep{lodders09}. For \nustar~ and joint \nustar~and \xmm~analyses, we used the abundance table of Wilms \citep{wilms00}. For the spectral analysis of \nustar\ data as well as for the joint fitting procedure of \nustar\ and \xmm\ data, photon counts used in spectra are grouped to have at least three counts per bin; therefore we do not apply the $\chi{^2}$ statistics which requires a minimum of $\sim$25 counts in each energy bin. 
Since the data follow a Poisson distribution, we applied the maximum likelihood-based statistic (hereafter, C-stat) appropriate for Poisson data as proposed by \citet{cash79}. However, with \xmm\ and \suzaku\ analyses, photon counts used in spectra are grouped to have at least 25 counts in each bin, and we used $\chi{^2}$ statistics to minimize differences in grouping and statistical approach with the literature. All uncertainties are quoted at the 68\% confidence level unless otherwise stated.

\section{Observations and data reduction} \label{sec:reduction}
In this work, we utilize \nustar, \xmm, and \suzaku~data to study the interface of A3395 and the intercluster filament that connects A3395 to A3391. The field of view (FOV) of the observations that are used in this work is overlaid on the recent \eros~events binned by 128~$\times$~128 as shown in Fig.~\ref{fig:erositaphoton}. We retrieved the \eros~data from the Early Data Release observations with Obs.ID: 300014\footnote{\url{https://erosita.mpe.mpg.de/edr/eROSITAObservations/}}.
The data from \xmm, and \suzaku~are retrieved from HEASARC archives\footnote{\url{https://heasarc.gsfc.nasa.gov/cgi-bin/W3Browse/w3browse.pl}}, whereas the \nustar~data were privately communicated to our group by the \nustar~SOC team as part of the \nustar~guest observer (Cycle-6) program award with an exclusive use period of one year. The public release date for the data was 2021-09-21. The specifications of the data are summarized in Table~\ref{tab:obslog}.

\subsection{\nustar} \label{sec:nustarbgd}

\nustar~observed the northwestern region of A3395 at the junction of cluster and intercluster filament (Fig.~\ref{fig:erositaphoton}) for $\sim$130 ks during 2020 with the focal plane modules (FPMA and FPMB) with an offset pointing from the central cluster emission (Table~\ref{tab:obslog}).

In order to filter the data, standard standard pipeline processing is carried out using HEASoft (v. 6.28) and NuSTARDAS (v. 2.0.0.) tools. Since the \nustar~observation of A3395 was performed after 2020 March 16, the use of versions of HEASoft earlier than v. 6.27.1 and of NuSTARDAS version earlier than v. 1.9.2 results in errors during pipeline processing\footnote{\url{http://nustarsoc.caltech.edu/NuSTAR_Public/NuSTAROperationSite/software_calibration.php/}}. To clean the event files, the stages 1 and 2 of the NuSTARDAS pipeline processing script nupipeline are used. Regarding the cleaning of the event files for the passages through the South Atlantic Anomaly (SAA) and a tentacle-like region of higher activity near part of the SAA, instead of using SAAMODE=STRICT and TENTACLE=yes calls, we have created light curves and applied different filters to create good time intervals (GTIs) manually without fully discarding the passage intervals.

\begin{figure*}
\centering
\includegraphics[width=159mm]{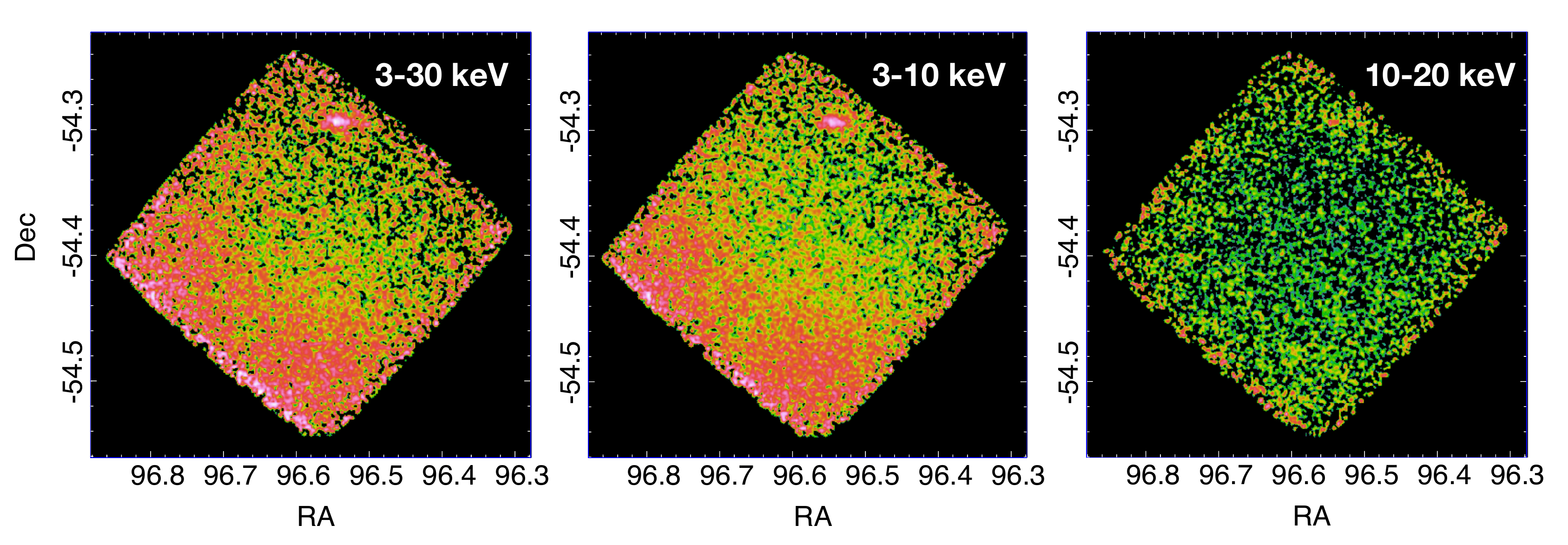}
\caption{\nustar~background subtracted, exposure corrected, smoothed photon images of A3395 at various energy bands. \label{fig:nuphoton}}
\end{figure*}

The new set of GTIs are reprocessed with nupipeline stages 1 and 2, and images are generated at different energy bands with XSELECT. nuexpomap is used to create exposure maps. To produce the corresponding spectra for the regions of interest as well as the corresponding response matrix files (RMFs) and ancillary response files (ARFs), stage 3 of nuproducts pipeline are used.

The background assessment of \nustar\ is particularly challenging due the open mast between the the focal plane modules and the optics assembly. The main components of the background are instrument Compton-scattered continuum emission, instrument activation and emission lines, cosmic X-ray background from the sky leaking past the aperture stops, reflected solar X-rays, and focused and ghost-ray cosmic X-ray background. Modeling the background where there is a lack of cluster emission regions is tricky since the ICM emission becomes an additional component in the background fitting procedure. 

To apply these models for the cluster background assessment, a set of IDL routines called ${\tt nuskybgd}$, which defines the background spatially and spectrally, is utilized \citep{wik14}. The procedure starts by selecting regions in the FOV where the cluster emission is inherently present yet not the most dominant. To account for the ICM emission, an apec model is included in the full set of models, and jointly fitted with the background (Fig.~\ref{fig:nuskybgd} in Section~\ref{sec:bgdfit}). The global background model is used to create background images, which are then subtracted from the cleaned images and are corrected by the corresponding exposure maps. Background-subtracted, exposure-corrected images at different energy bands are presented in Fig.~\ref{fig:nuphoton}.

Once the background is defined for any region in the FOV both spatially and spectrally, the next steps are to select regions of interest, to extract spectra and the corresponding files, and to generate the specific background model, followed by spectral fitting to evaluate the physical properties of the ICM.

\subsection{\xmm}

\xmm~observations lasted for $\sim$30 ks during 2007 (Table~\ref{tab:obslog}). The FOV of this observation is focused on the central region of A3395, which extends to $\sim$r$_{500}$ (Fig.~\ref{fig:erositaphoton}). The three EPIC cameras MOS1, MOS2 \citep{turner01}, and PN \citep{struder01} were operated in full frame mode with the Thin1 filter.

We used standard procedures from the Science Analysis System (SAS) software version 16.1.0, along with the Extended Source Analysis Software (ESAS\footnote{\url{https://heasarc.gsfc.nasa.gov/docs/xmm/esas/cookbook/xmm-esas.html}}) package. Calibrated photon event files were produced using the {\bf epchain} and {\bf emchain} tasks. For obtaining the good time intervals, we used {\bf mos-filter} and {\bf pn-filter} tasks. Point sources were detected by the {\bf cheese} routine.

ESAS routines {\bf mos\_back} and {\bf pn\_back} embedded in SAS create model quiescent particle background (QPB) spectra and images for selected regions from the intermediate files produced from the {\bf mos-spectra} and {\bf pn-spectra}. To create background-subtracted, exposure-corrected images, first a spectrum for the source and the QPB was extracted from the full FOV. This spectrum was fitted with the absorbed apec model for the cluster emission plus the models for the background. The background model components are the cosmic diffuse X-ray background (CXB), solar wind charge exchange (SWCX), particle and residual soft proton contamination (K$\alpha$~SP). 
\begin{figure}
\centering
\includegraphics[width=86mm]{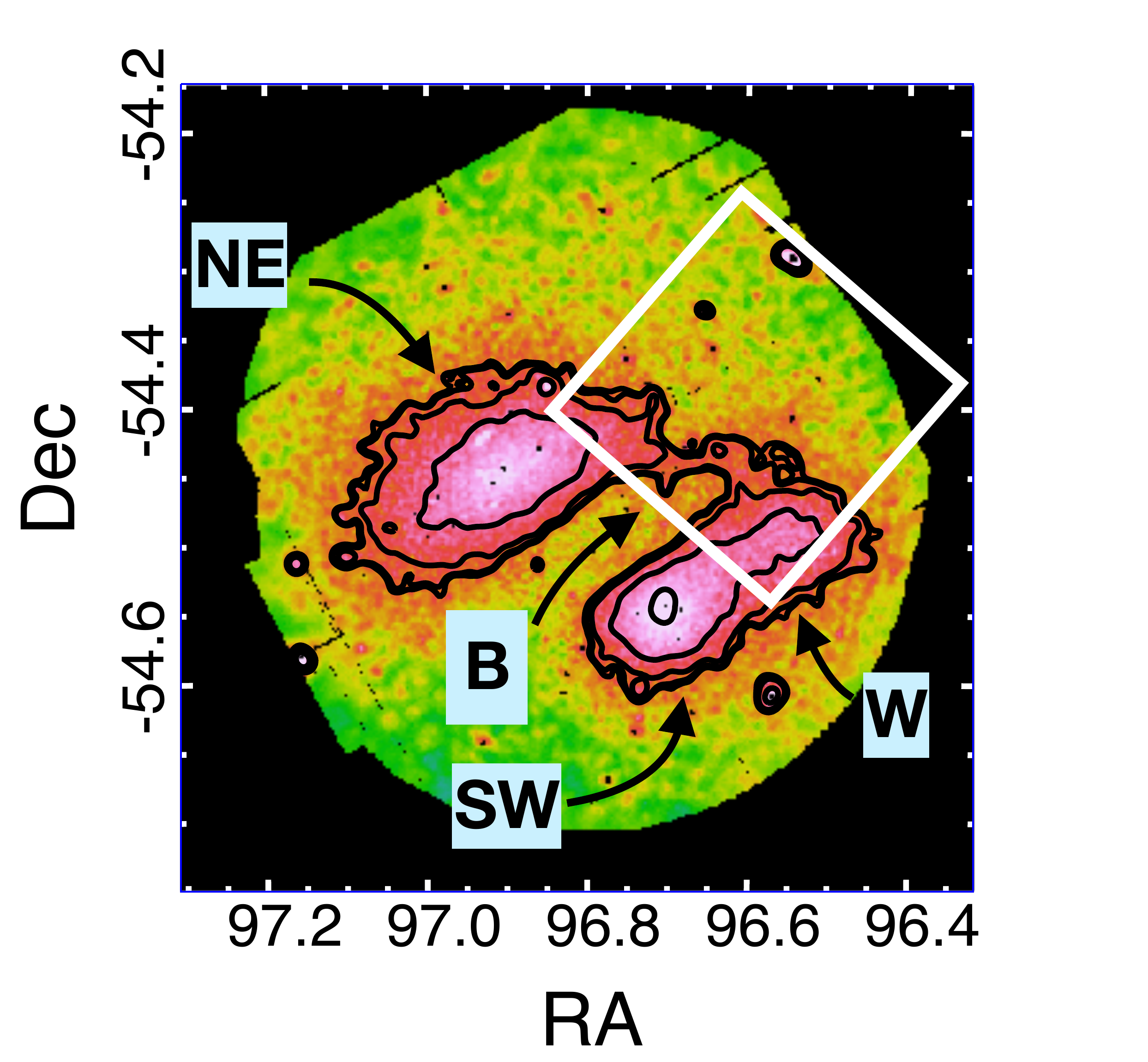}
\caption{Background subtracted, exposure corrected, adaptively smoothed \xmm\ photon image in the soft X-ray (0.4-2.5 keV) band. \nustar\ field of view is indicated with the white box region. NE and SW are subclusters of A3395, B indicates the bridge emission in between the subclusters, and W indicates the Western subclump as described in Section~\ref{sec:intro}.\label{fig:xmmfigure}}
\end{figure}

CXB is modeled with an unabsorbed thermal component of about 0.1 keV, an absorbed thermal component of about 0.25 keV, and the extragalactic power law with a spectral index of 1.46. The particle background is Gaussian lines at 1.496 keV and 1.75 keV representing the Al K$\alpha$~ and Si K$\alpha$~ lines in the MOS, lines at 1.496 and near 8 keV representing the Al K$\alpha$~ and Cu fluorescent lines in the PN. Possible SWCX produces two more lines at 0.56 and 0.65 keV, and the residual SP is represented by a broken power law and fitted with a diagonal matrix supplied in the XMM-ESAS CalDB release, mos1-diag.rsp.gz, mos2-diag.rsp.gz, and pn-diag.rsp.gz. 
To find the solid angle on each detector for the fit {\bf proton\_~scale} was used, which were used as a {\bf constant} model between the cameras in the fit model. The routine {\bf proton} takes the fitted parameters from the fit of the SP model, then creates SP background images later to be subtracted as a background from the photon image.

This process is repeated for an annulus selected in an outer region of the FOV to exclude the bright cluster emission at the center of the FOV in order to produce a better fit for the SP contamination. The resulting SP normalization from the annulus fit is then renormalized to the full FOV using {\bf proton\_scale}. The resulting adaptively smoothed, background-subtracted, exposure-corrected images are obtained with tasks {\bf comb} and {\bf adapt} respectively, as shown in Fig.~\ref{fig:xmmfigure}.

In addition to the ESAS background, the total background model used to fit the spectra contains the following components: {\bf gaussian} + {\bf gaussian} + {\bf gaussian} + {\bf gaussian} + {\bf gaussian} + {\bf gaussian} + {\bf gaussian} + {\bf constant}$\times${\bf constant}$\times$({\bf gaussian} + {\bf gaussian} + {\bf apec} + ({\bf apec} + {\bf apec} + {\bf powerlaw}$\times${\bf wabs}), where the second {\bf constant} corresponds to the cross calibration of instruments MOS1, MOS2 and PN. This background modeling is used for the rest of the spectral analysis in this paper.

\subsection{{\it Suzaku}}
{\it Suzaku} \citep{mitsuda07} also covered a region enclosing the \nustar~FOV (Fig.~\ref{fig:erositaphoton}) for $\sim$47 ks in 2013.
The X-ray Imaging Spectrometer \citep[XIS:][]{koyama07}) of Suzaku is one of the best instruments to investigate shallow cluster emission in the outskirts owing to their low and stable background. The basic data analysis and results are presented in \citet{sugawara17}. Here we briefly explain the data reduction and spectral analysis approach. 
\\
We followed the approach presented in \citet{sugawara17}. We used the latest calibration file (20160607) and performed event screening with cutoff rigidity greater than 8 GV. The area damaged by micrometeoroids in the XIS0 instrument was excluded in the following analysis (\url{http://www.astro.isas.jaxa.jp/suzaku/doc/suzakumemo/suzakumemo-2010-01.pdf}). An additional screening is applied for XIS1 to mitigate the increase in non-X-ray background (NXB) due to an increase in the amount of charge injection (\url{http://www.astro.isas.jaxa.jp/suzaku/analysis/xis/xis1_ci_6_nxb}). 
The cleaned exposure time is about 33 ks for each instrument. 
%%%
In order to characterize the ICM emission, all background components need to be constrained well. 
For the estimation of the NXB, we used a database constructed from observations of Earth at night using the ftool ${\tt xisnxbgen}$ \citep{tawa08}. 
For the sky background consisting of cosmic-ray background, local hot bubble, and Milky Way halo, we used same model presented in Table 2 of \citet{sugawara17}.
\\
RMFs and ARFs are generated by the ftools ${\tt xisrmfgen}$ and ${\tt xissimarfgen}$ \citep{ishisaki07}. For the ARFs, we assumed uniform emission from a circular region with 20$\arcmin$ radius as an input image. Throughout the fitting procedure, the normalization between BI and FI instruments is kept free.

\section{Data analysis and results} \label{sec:analysis}
\subsection{Global view} 

We start the analysis with the characterization of the global emission at the \nustar~FOV that is pointed at the northwestern region of A3395. We selected a square 12$\arcmin$ $\times$ 12$\arcmin$ region from which we extracted a spectrum and applied a single-temperature ${\tt apec}$ model \citep{smith01} using ${\tt XSPEC}$ \citep[v.~12.11.1;][]{arnaud96}. Since \nustar\ is not sensitive to emission below 3 keV, it is also not affected by foreground absorption by the Galactic hydrogen column density, and any changes in the $N_H$ value do not affect the thermodynamical values obtained from the fit. We visually inspected the images for the point sources, and excluded a 1$\arcmin$ (comparable to the half-power diameter of \nustar's point spread function) circular region from the location of the point source that is visible in Fig.~\ref{fig:nuphoton}.

\begin{figure}
\includegraphics[width=83mm]{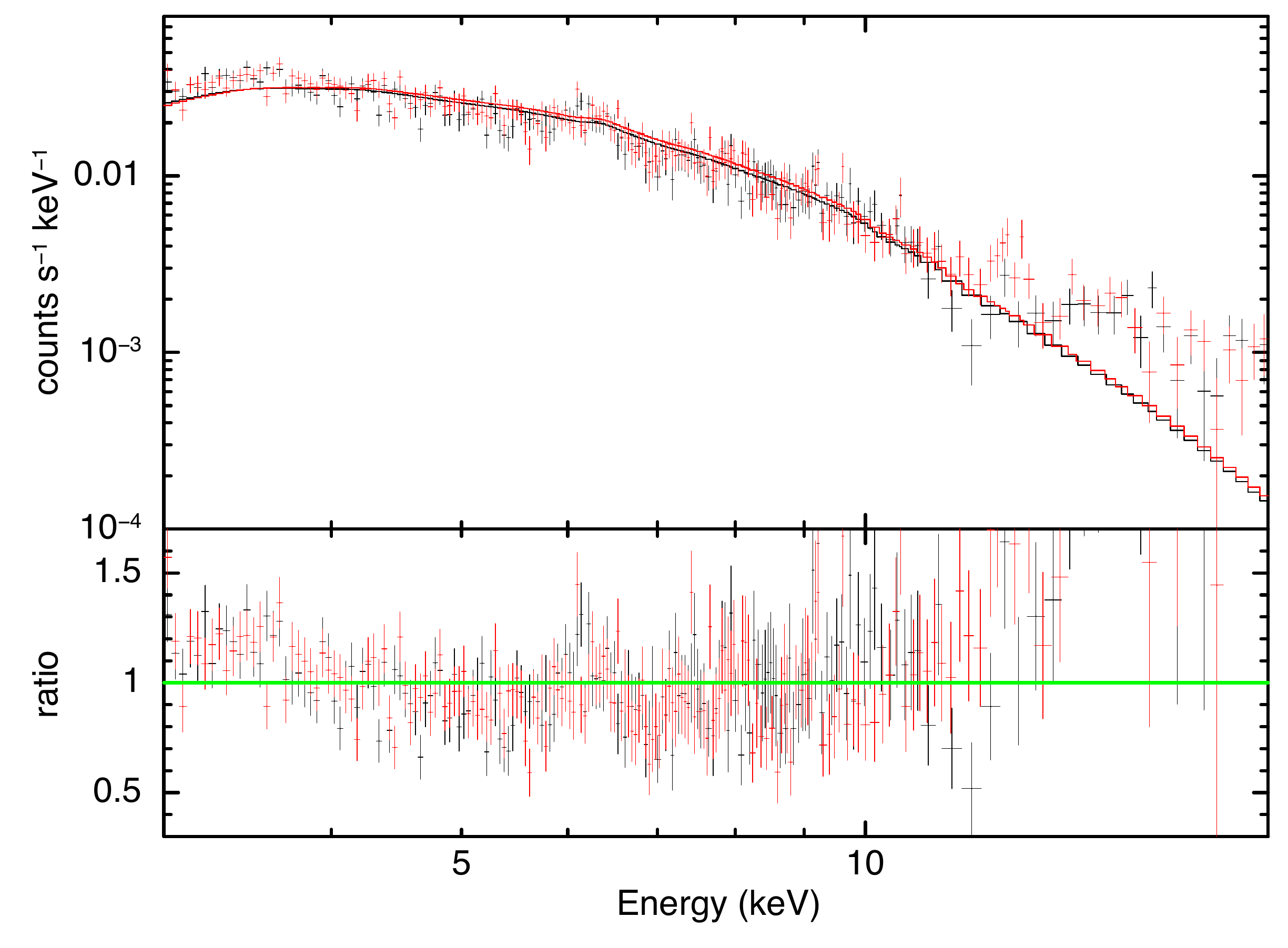}
\includegraphics[width=83mm]{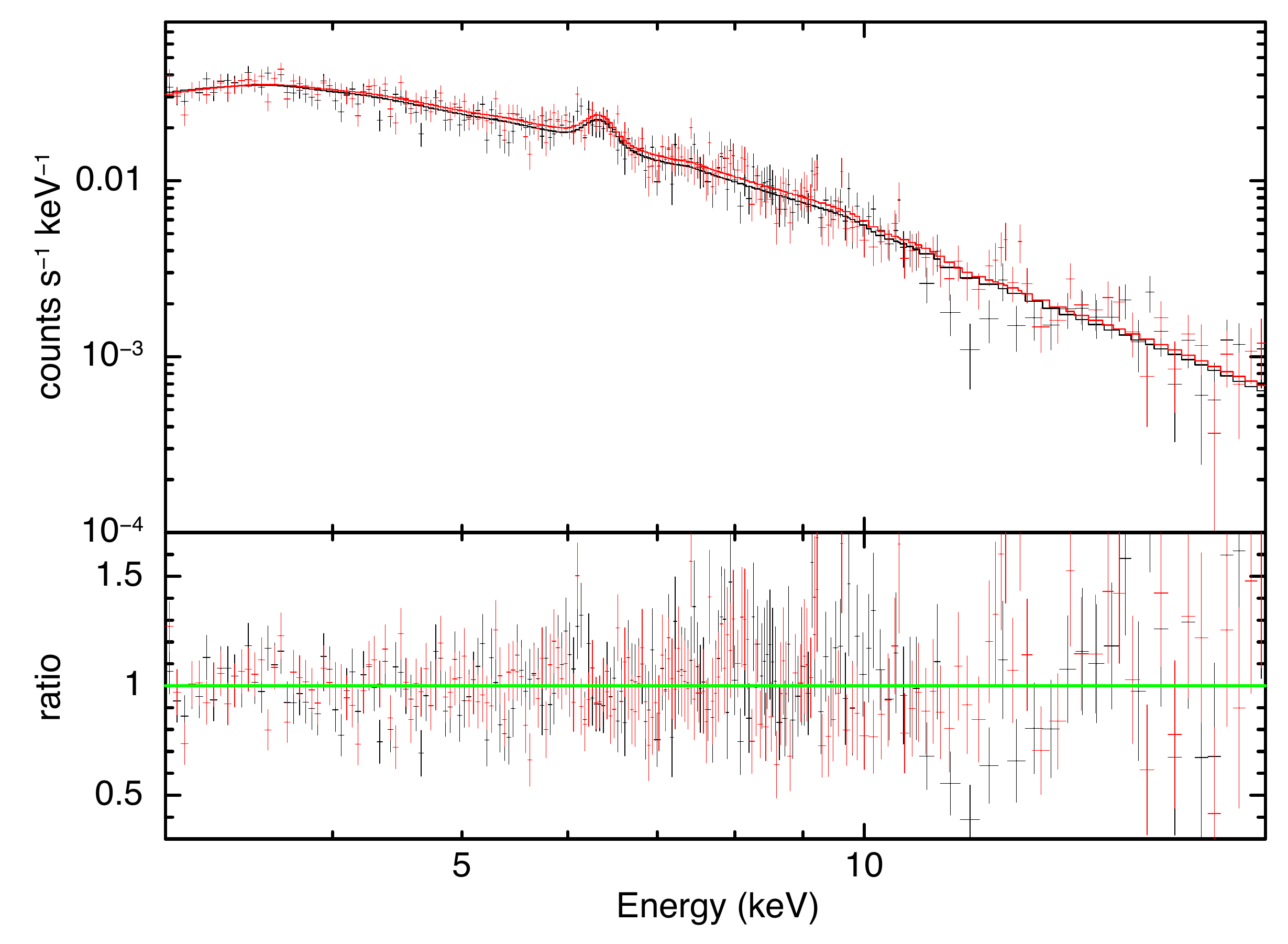}
\caption{Global fits with single {\bf apec} ({\it upper panel}) and {\bf apec} + {\bf powerlaw} ({\it lower panel}) of the $12\arcmin \times 12\arcmin$ region. \label{fig:global}}
\end{figure}

The spectral fit and the corresponding values are presented in Fig.~\ref{fig:global} and Table~\ref{tab:globalfit}, respectively. It is evident from the spectra and the C-stat values that a single-temperature plasma does not fully describe the cluster. This is expected since the \nustar\ observation is pointed at a region where there are multiple substructures, which may involve emission features from different sources. To assess the possibility of multiple temperatures, we have added another ${\tt apec}$ model for a secondary temperature structure, yet the higher-temperature component was not constrained.

One of these emission features can also be due to a nonthermal component, namely inverse Compton (IC) scattering, which is well represented by a power-law distribution and corresponds to the model ${\tt powerlaw}$ in ${\tt XSPEC}$. The addition of a ${\tt powerlaw}$ model to the original ${\tt apec}$ seems to describe the overall cluster emission well, but the reason for this extra component may be completely unrelated to cluster physics, namely scattered light caused by the two main subclusters lying just outside the FOV. This additional component also seems to suppress the ICM to an unrealistic low temperature with respect to what is found in the literature \citep{markevitch98,lakhchaura11,alvarez18}, further suggesting a scattered light origin.

\begin{deluxetable*}{lcccccc}
\tablecaption{Spectral parameters of \nustar\ for the global analysis. ${\tt apec}$ normalization ({\it norm}) is given in $\frac{10^{-14}}{4\pi \left [ D_A(1+z) \right ]^{2}}\int n_{e}n_{H}dV$  where ${\tt powerlaw}$ normalization ($\kappa$) is {\it photons}  keV$^{-1}$ cm$^{-2}$ s$^{-1}$ at 1 keV. All errors are quoted at 68\% confidence level.
\label{tab:globalfit}}
\tablehead{\\[-0.95em]
& \colhead{{\it kT}}& \colhead{{\it Z}} & \colhead{{\it norm}} & \colhead{$\Gamma$} & \colhead{{\it $\kappa$}} & \colhead{{\it C / $\nu$}} \\
[-0.95em]
& \colhead{(keV)}& \colhead{({\it Z$_{\odot}$})} & \colhead{(10$^{-2}$ cm$^{-5}$)} &  & \colhead{(10$^{-3}$)} & \colhead{cstat/d.o.f}
}
\startdata
\\[-0.95em]
${\tt apec}$ & 5.59 $\pm{0.11}$ & 0.05 $\pm{0.02}$  & 1.42 $\pm{0.03}$ & \nodata & \nodata & 1219.22/844\\
\\[-0.95em]
${\tt apec}$ + ${\tt powerlaw}$ & 2.06$^{+0.31}_{-0.22}$ & 0.36$^{+0.16}_{-0.11}$ & 2.13$^{+0.19}_{-0.23}$ & 1.82$^{+0.18}_{-0.29}$ & 1.75$^{+1.04}_{-0.95}$ & 902.55/842\\
\\[-0.95em]
\enddata
\end{deluxetable*}

\subsection{6 x 6 grid analysis} \label{sec:grid}

\begin{figure}
\centering
\includegraphics[width=83mm]{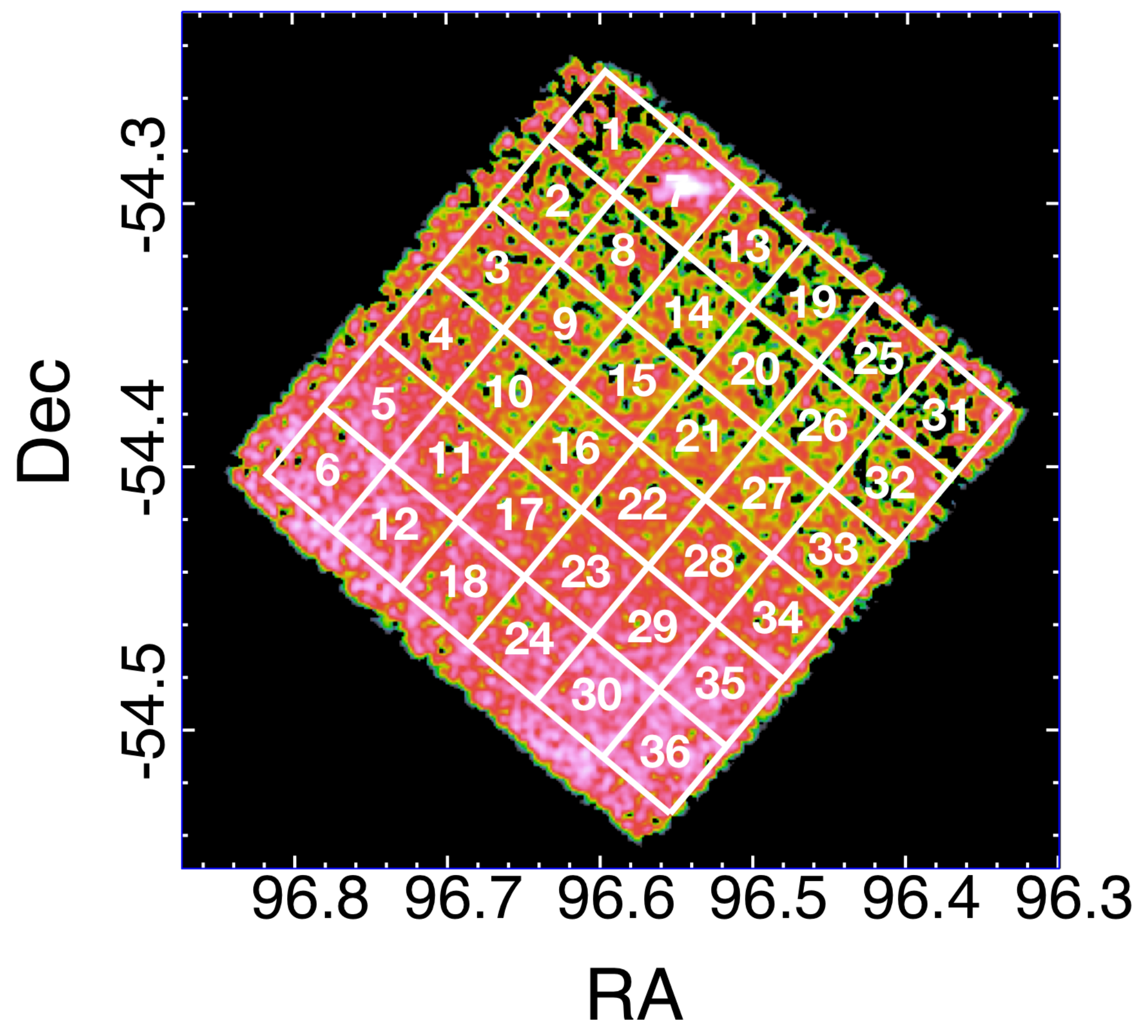}
\caption{Grid system and region numbers superimposed on background subtracted, exposure corrected \nustar~photon images of A3395 in 3-20 keV energy band. The spectrum of each region is fitted with a single temperature model to build the thermodynamical profiles. \label{fig:grid}}
\end{figure}

Since A3395 has many substructures within the \nustar\ FOV, a typical assumption of spherical symmetry cannot be used for region selection. For the characterization of the regions, we have created thermodynamical maps of a 12$\arcmin$ $\times$ 12$\arcmin$ square region, encompassing a 6 $\times$ 6 grid system as shown in Fig.~\ref{fig:grid}. This method is used to aide our understanding of the cluster ICM in detail, as done by \citet{gastaldello15} for the \nustar\ observation of the Coma cluster. We use these results to select regions with similar thermodynamical properties to define larger regions with higher signal-to-noise ration (S/N). 

For this analysis, we calculated projected maps, since the main goal is to achieve a comparative study, rather than to study the specific density, entropy, and pressure properties. To assess the ICM properties of box 7, we excluded the point source in that region. Although there will also be scattered light contamination in these regions, we expect the scattered light to change mildly in the adjacent regions due to the small region size we select. This means that, even if a most complete assessment of ICM properties may not be managed, we aim to still have a good proxy for the definition of regions of interest.

\begin{figure}
\includegraphics[width=86mm]{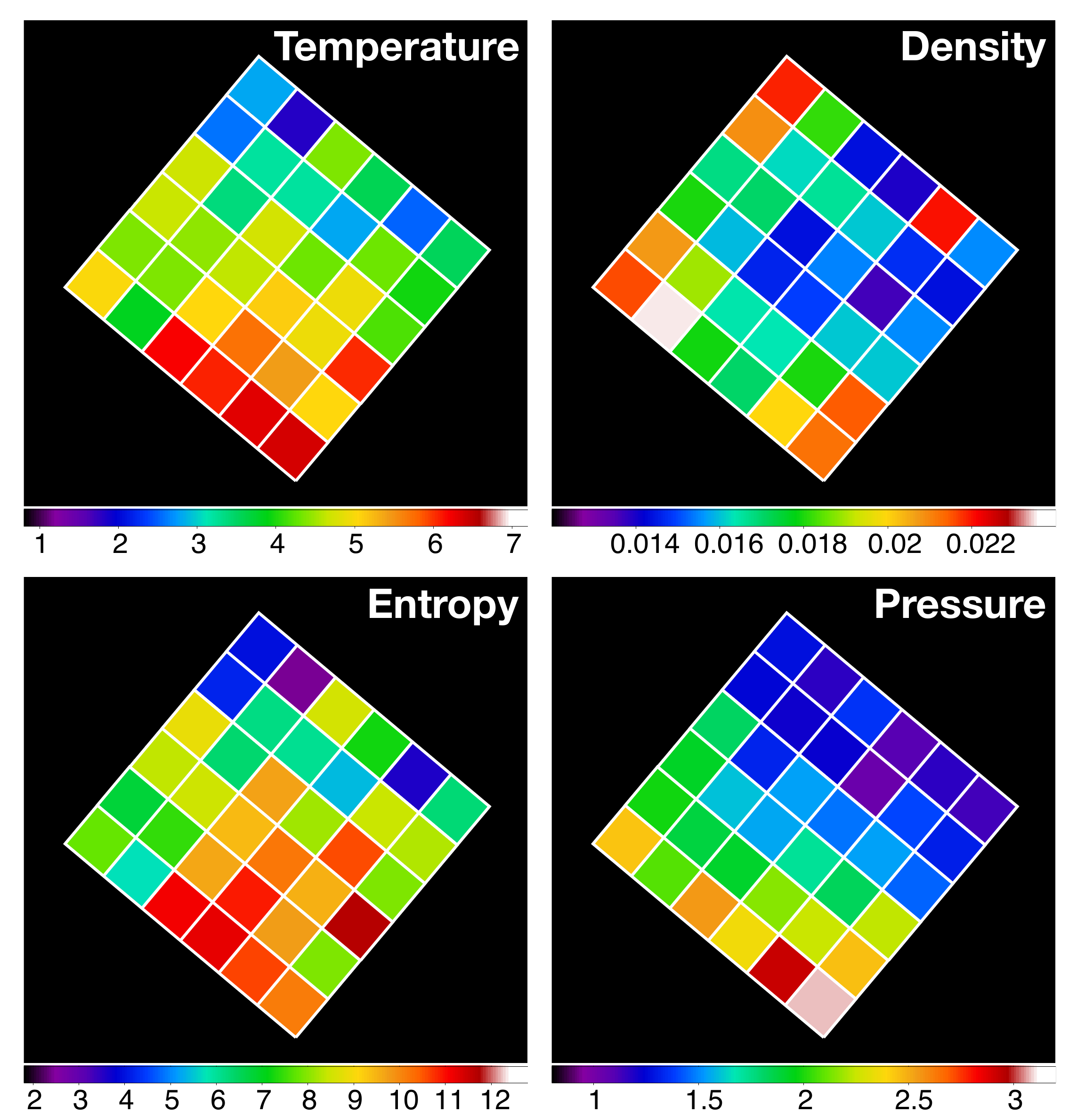}
\caption{Projected \nustar~temperature ({\it upper left}), projected density ({\it upper right}), pseudo-entropy ({\it lower left}) and pseudo-pressure maps ({\it lower right}). Temperature units are given in keV, whereas the rest of the maps are presented in arbitrary units and do not account for the possible scattered light contamination.
\label{fig:gridmap}}
\end{figure}

\begin{figure*}
\centering
{\includegraphics[width=183mm]{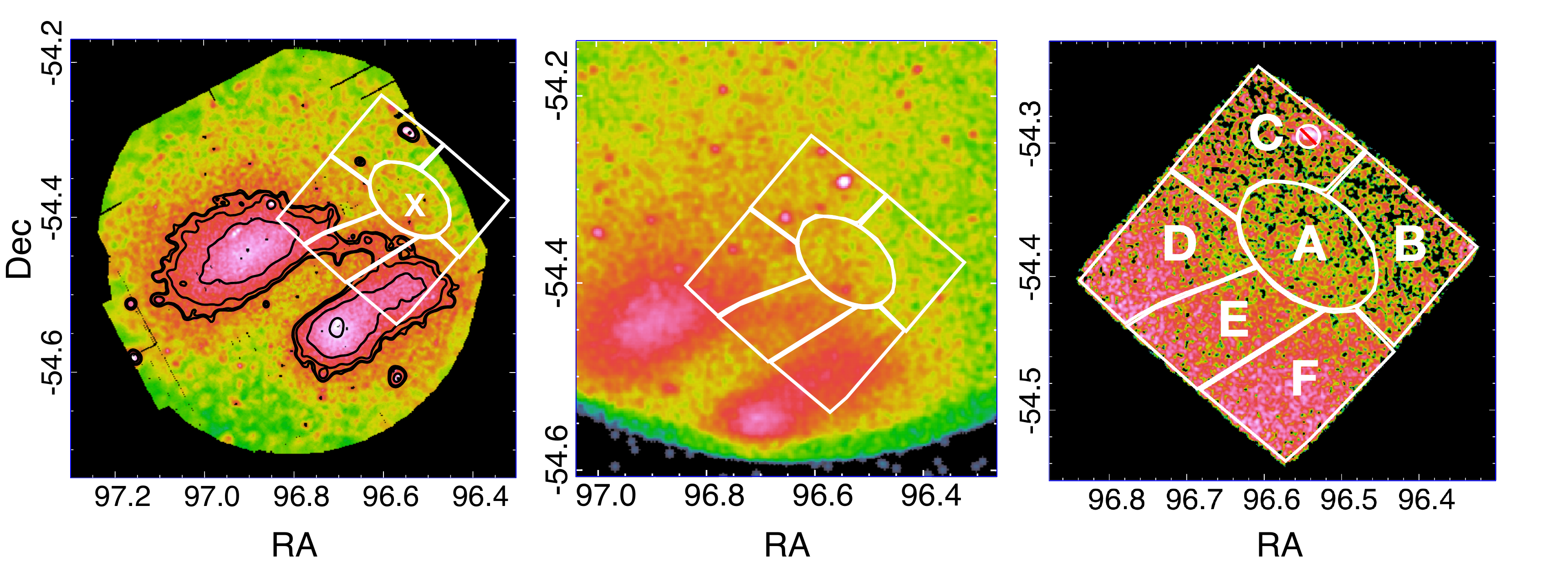}}
\caption{Regions of interest superimposed on 0.4-2.5 keV \xmm~({\it left panel}), on \eros~({\it middle panel}) and 3.0-20.0 keV \nustar~({\it right panel}) photon images of A3395. X label on the \xmm~ image indicated the location of the hot spot found in literature \citep{lakhchaura11}.\label{fig:6reg}}
\end{figure*}

The grid system analysis is achieved with \nustar\ data, where we extracted spectra from each grid then fitted them using a single-temperature ${\tt apec}$ model. The abundances for the fits are fixed to {\it Z}=0.3 {\it Z$_{\odot}$}, because they were difficult to constrain due to low S/N. For the selection of regions with similar thermodynamical properties for the whole FOV, maps of \nustar\ temperature, density, entropy, and pressure were created. The normalization of the ${\tt XSPEC}$ model ${\tt apec}$ is defined as $\left [10^{-14}/4\pi \left [D_{A}(1+z) \right ]^{2} \right ]\int n_{e}n_{H}dV$, where the integrand is the emission measure (EM), n$_{e}$ and n$_{H}$ are electron and hydrogen densities in cm$^{-3}$, and the angular diameter distance to A3395 is D$_{A}$ $\simeq$ 6.20 $\times$ 10$^{26}$ cm. By using the normalization of ${\tt apec}$ model, we estimated the EM, and the corresponding pseudo-pressure and pseudo-entropy maps using {\it P} = {\it kT} $\times$ EM$^{1/2}$ and {\it S} = {\it kT}/EM$^{1/3}$ \citep[e.g.][]{rossetti07}. The projected density is calculated as the square root of the normalization parameter of the ${\tt apec}$ model.

Following the results of our grid analysis and with the guidance of the \xmm~photon image, we defined 6 regions of interest on the \nustar\ FOV Fig.~\ref{fig:6reg}. Region A represents the location of the hot spot detected by \xmm~ \citep{lakhchaura11}, Region B is the region extending to the intercluster filament, Region E is the bridge between the subclusters of A3395, and Region F is the tail of the southwestern subcluster (or the W clump). Region C is isolated since it shows higher-density regions than Regions A and B from our \nustar~grid analysis (Fig.~\ref{fig:gridmap}), and also shows excess emission in the \eros~image, probably of filamentary origin (Fig.~\ref{fig:erositaphoton} and Fig.~\ref{fig:6reg}, middle panel). We note that Regions A, B, and C are mostly enclosed in the region isolated by \citet{lakhchaura11} (indicated by NW), which they suggest that connects the cluster to the intercluster filament.

\subsection{Scattered light and ray-trace simulations} \label{sec:SL}

\nustar\ is affected by scattered light contamination \citep[also known as Ghost Rays;][]{madsen17} due to X-ray photons that undergo only a single reflection off of either the primary (upper) or secondary (lower) mirror, as opposed to a properly focused double reflection. Due to the lack of precollimators, photons from bright sources outside the FOV can reach the focal plane without double reflection.

The upper single reflection is due to the photons that strike the upper mirrors at angles steeper than the nominal focusing graze angle; it fades away when the angle becomes too steep so that the adjacent shell shadows it \citep{madsen17}. The aperture stop also helps to block some of these photons. The lower single reflection, on the other hand, is caused by photons reaching the mirrors at angles that are shallower than the nominal graze angle \citep{madsen17}. In addition, back reflections can occur, in which photons hit the back side of the upper mirror of the adjacent shell first, then reflect off the front side of the mirror shell. The scattered light contamination appears as early as 2$\arcmin$ off-axis but only becomes significant above 3$\arcmin$, out to $\sim$1$\arcdeg$ \citep{madsen17}. The off-axis and energy dependence of scattered light is understood for point sources \citep{madsen17}, but no data analysis tool yet exists to model its effect in the case of extended emission. Ray-trace simulators, which trace out the paths of photons through the optics and onto the detectors, have been shown to reproduce observed scattered light patterns \citep{westergaard11,madsen17}.

We analyzed the effect of the scattered light for the \nustar\ observation, in an attempt to explain the excess power-law emission in the global spectra. Also, it is already expected since the two main subclusters, namely NE and SW (Fig.~\ref{fig:6reg}), lie near the FOV of the \nustar\ pointing. Our FOV covers distances of 3$\arcmin$-20$\arcmin$ from either of the subcluster centers, where the scattered light contamination is known to be significant \citep{madsen17}.

We did not study the effect of the back reflection contamination, since this component appear at very high fluxes and becomes effective at $\sim$15$\arcmin$, whereas the largest distance covered by our \nustar\ FOV from either of the subcluster centers is $\sim$20$\arcmin$. Also, the relative geometric area, i.e., fraction of back reflection to on-axis geometric area, is below 1\% between 15$\arcmin$ and 20$\arcmin$ \citep{madsen17}.

To model the scattered light, we used the ray-tracing code for X-ray telescopes, MT\_RAYOR (version 4.6.9) \citep{westergaard11} written in Yorick interpreted language \citep{munro95}, to simulate the properly focused photons (good photons) and photons that undergo single reflection (scattered light). MT\_RAYOR takes a count rate image in the 0.5-2.5 keV energy band and a temperature map constructed from other X-ray missions---\xmm\ in this case---guided by literature \citep{lakhchaura11}, as an input for the distribution of photons in the cluster, divides the cluster into predefined pixels, and predicts the expected double and single bounce photons from a \nustar\ observation, mainly treating the extended ICM emission as a collection of point sources that have different spectral properties as well as photon distribution.

\begin{figure}
\centering
\includegraphics[width=84mm]{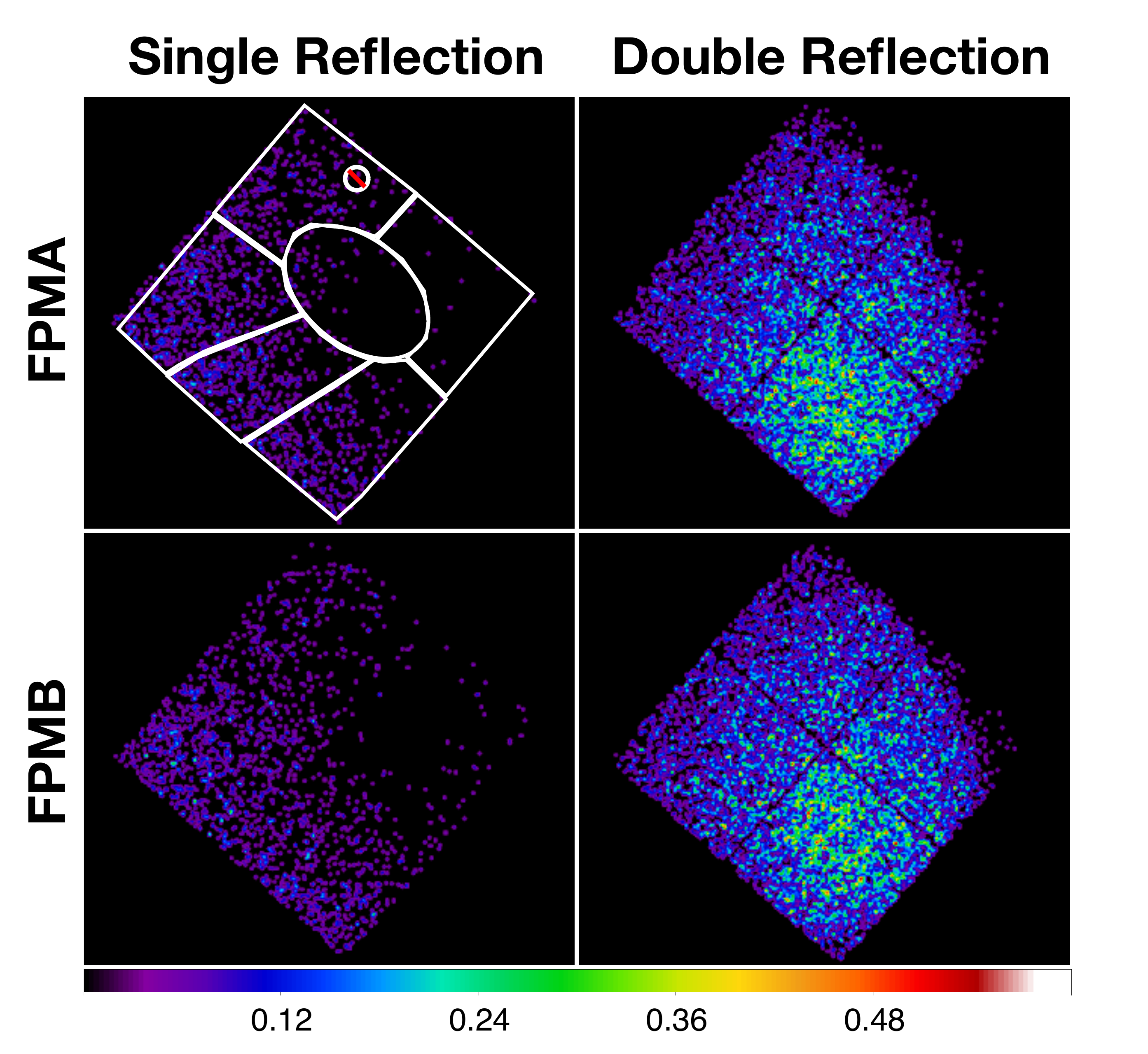}
\caption{Ray-trace simulation results obtained from MT\_RAYOR indicating single reflection photons (scattered light, {\it left panel}), and photons that are properly focused ({\it right panel}) on the focal plane modules A (FPMA, {\it top panel}) and B (FPMB, {\it bottom panel}). Regions of interest are overlaid on FPMA scattered light photon distribution. \label{fig:scatter}}
\end{figure}

These MT\_RAYOR simulations provide a spatial distribution of good photons as well as the scattered light. We then extracted spectra from the 6 regions from both the good photons and scattered light simulation results (Fig.~\ref{fig:scatter}). Good photons describing the cluster emission were then fit by a single ${\tt apec}$ model, whereas we applied different models to fit the scattered light to find the best model that describes this contamination. We used ${\tt apec}$, ${\tt powerlaw}$, and ${\tt bknpower}$ models and find that ${\tt powerlaw}$ is the best model describing the scattered light for all regions. The model ${\tt powerlaw}$ has two parameters, photon index ($\Gamma$) and model normalization ($\kappa$). The resulting photon indices are presented in Table~\ref{tab:simfitresults}. The spectral fits of the simulated regions are shown in Fig.~\ref{fig:scatterfit} in Section~\ref{sec:raytracefit}.

\begin{figure}
\centering
\caption{Scattered light recipe. \label{fig:recipe}}
\vskip 0.3 cm
\begin{tikzpicture}[node distance = 1.8cm, auto]
\node [block] (xmm) {Create \xmm\ temperature map using literature, and archival data to create a photon image as an input for ray-tracing};
\node [block, below of=xmm] (init) {Simulate scattered photons and good photons using MT\_RAYOR ray-trace simulations};
\node [block, below of=init] (spectra) {Extract ${\tt XSPEC}$ compatible spectra from 6 ROI for scattered photon and good photon simulations separately};
\node [block, below of=spectra] (fit) {Fit the scattered spectra with various models and good photons with ${\tt apec}$ model from 6 ROI};
\node [block, below of=fit] (identify) {Identify best fit model and related parameters};
\node [block, below of=identify] (calculate) {Calculate the flux ratio of scattered light to total photons from 6 ROI using MT\_RAYOR ray-trace simulations};
\node [block, below of=calculate] (dataflux) {Calculate total data flux for 6 ROI from \nustar\ observational data};
\node [block, below of=dataflux] (fakeit) {Use ${\tt fakeit}$ to simulate scattered light spectra};
\node [block, below of=fakeit] (addspec) {Use ${\tt addspec}$ to add the corresponding ${\tt fakeit}$ simulated spectra to each 6 ROI background spectra obtained from ${\tt nuskybgd}$};
    \path [line] (xmm) -- (init);
    \path [line] (init) -- (spectra);
    \path [line] (spectra) -- (fit);
    \path [line] (fit) -- (identify);
    \path [line] (identify) -- (calculate);
    \path [line] (calculate) -- (dataflux);
    \path [line] (dataflux) -- (fakeit);
    \path [line] (fakeit) -- (addspec);
\end{tikzpicture}
\end{figure}
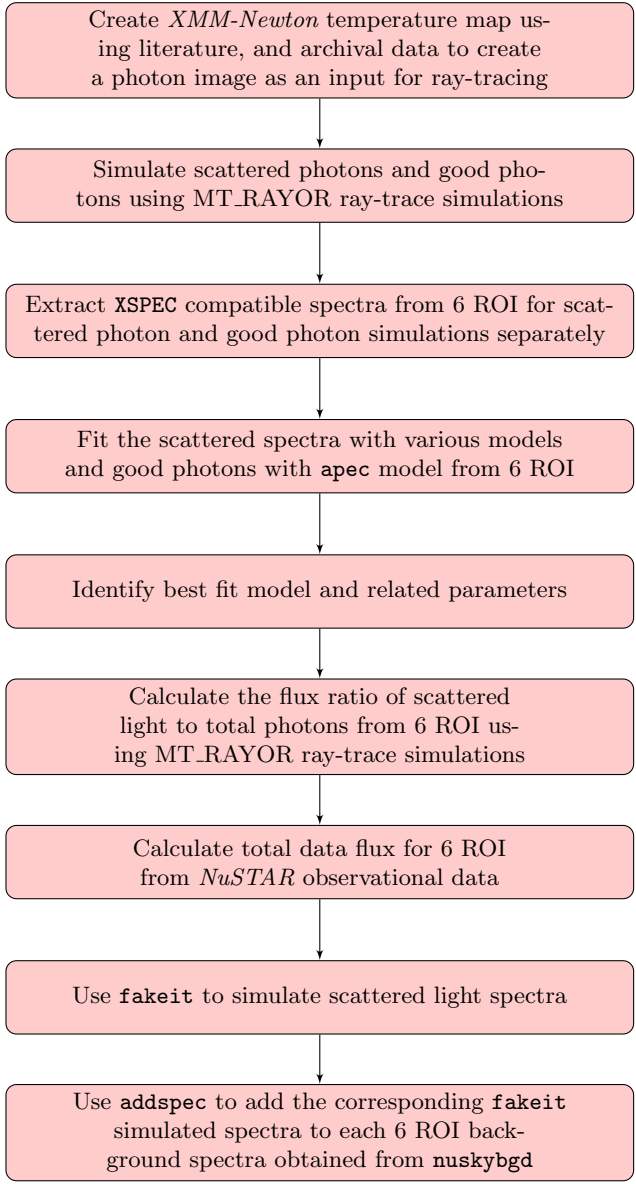

We also created hardness ratio maps with a selection of different energy bands to investigate the energy dependence of the spatial distribution of the ray-traced scattered light, yet no apparent gradient was observed.

\begin{deluxetable*}{lcccccc}
\tabletypesize{\scriptsize}
\tablewidth{0pt} 
\tablecaption{Photon index ($\Gamma$) values from ${\tt powerlaw}$ fit of ray-traced \nustar\ scattered light for the regions of interest in energy band 3-15 keV. The scattered light flux percentages with respect to total photons are given inside parenthesis.
\label{tab:simfitresults}}
\tablehead{\\[-0.95em]
& \colhead{Region A} & \colhead{Region B} & \colhead{Region C} & \colhead{Region D} & \colhead{Region E} & \colhead{Region F}}
\startdata
FPMA & 2.44 (1.98$\%$) & N/A ($<$1$\%$) & 2.64 (11.0$\%$) & 2.28 (23.3$\%$) & 2.31 (12.4$\%$) & 2.45 (10.4$\%$) \\  
\\[-0.95em]
FPMB & 3.50 (2.21$\%$) & 2.01 (4.19$\%$)  & 2.40 (10.8$\%$) & 2.53 (24.0$\%$) & 2.67 (14.8$\%$) & 2.14 (12.9$\%$)\\
\\[-0.95em]
\enddata
\end{deluxetable*}

Once we calculated the 3-15 keV flux from both the good photons and scattered light for all six regions of interest, we deduced the ratio of the scattered photons to the total photons as presented in Table~\ref{tab:simfitresults}. The flux ratios as well as simulated images showed that the main region of interest, i.e. Region A, showed very little contamination from the scattered light. Since the number of ray-trace-simulated photons may differ from the real data, direct addition of the ray-traced scattered light spectra to the background spectrum may result in overestimation or underestimation of scattered light. Therefore, we calculated the fluxes from all six regions of the observational data, then with the 3-15 keV flux ratios provided from the aforementioned analysis, we used the ${\tt fakeit}$ function of ${\tt XSPEC}$, to simulate the scattered light with the previously obtained fitted parameter $\Gamma$, and reset the model norms to match the flux-estimated ray-trace flux ratios. Our recipe is summarized in the flow chart in Fig.~\ref{fig:recipe}.

\subsection{The \nustar, \xmm, \suzaku, and joint \nustar\ and \xmm\ analysis of regions of interest}\label{sec:allfit}

We fitted the ${\tt apec}$ model to the \xmm\ spectrum extracted from the six ROI. In this analysis, the abundance was a free parameter, and we present best-fit results where the {\it N$_H$} is  fixed to the literature value of {\it N$_H$} = 6.30 $\times$ $10^{20}$ cm$^{-2}$ \citep{lakhchaura11}, in Table~\ref{tab:allfit}. 

The faked spectra of each region were added to the corresponding background model spectra using ${\tt addspec}$ script by HEAsoft ${\tt ftools}$. After combining the scattered light and ${\tt nuskybgd}$ background, we fit the \nustar\ observational data with an ${\tt apec}$ model for the six regions, and the results are shown in Table~\ref{tab:allfit}. 

For comparison and to see the effect of scattered light on model parameters, we also repeated the analysis without including our scattered light analysis. The results for this secondary analysis are shown in Table~\ref{tab:allfit}.

We also repeated the spectral analysis for the same six ROI using \suzaku\ data. The result of the fits are presented in Table~\ref{tab:allfit}.

As the next step, we fitted \nustar\ and \xmm\ spectra jointly for all 6 ROI, where the abundance values were fixed to the values obtained from \xmm\ spectral fits. The results are shown in Table~\ref{tab:allfit}.

We also jointly fitted the \nustar\ and \xmm\ spectra from the regions of interest using an additional ${\tt apec}$ or ${\tt powerlaw}$ component to the original ${\tt apec}$ model. From these regions, the emission coming from Regions A and E is found to be better described with an additional spectral model. 

Region A hints at two-temperature structure, the high-temperature component being {\it kT} = 16.2 keV, for which only a lower limit of 8.5 keV is obtained. The lower temperature is found to be {\it kT} = 2.90$^{+0.47}_{-0.29}$ keV. This ${\tt apec}$ + ${\tt apec}$ model, improved the statistics by {\it $\Delta$C/$\Delta\nu$} = 12.47/2. Applying ${\tt apec}$ + ${\tt powerlaw}$ model also improved the statistics ({\it $\Delta$C/$\Delta\nu$} = 12.36/2), giving a photon index of $\Gamma$ = 1.70$^{+0.23}_{-0.62}$.

\begin{deluxetable*}{lcccccc}
\tabletypesize{\scriptsize}
\tablewidth{0pt} 
\tablecaption{Spectral parameters of \xmm~(0.5-9.0 keV), \nustar~(3.0-15.0 keV), and \suzaku~(0.7-7.0 keV) analysis for the regions of interest. ${\tt apec}$ normalization (${\it norm}$) is given in $\frac{10^{-14}}{4\pi \left [ D_A(1+z) \right ]^2}\int n_{e}n_{H}dV$.
\label{tab:allfit}}
\tablehead{\\[-0.95em]
\colhead{Regions}& \colhead{Spectral} & \colhead{\xmm~\tablenotemark{a}} & \colhead{\nustar~} & \colhead{\nustar} & \colhead{\suzaku~\tablenotemark{a}} & \colhead{Joint \nustar~\tablenotemark{b}}\\[-0.95em]
\colhead{}& \colhead{Parameters} & \colhead{} & \colhead{with} & \colhead{without} & \colhead{} & \colhead{and}\\[-0.95em]
\colhead{}& \colhead{} & \colhead{} & \colhead{SL treatment} & \colhead{SL treatment} & \colhead{} & \colhead{\xmm}}
\startdata
\\[-0.95em]
{A} & {\it kT} (keV) & 4.46$^{+0.78}_{-0.49}$ & 3.74$^{+0.40}_{-0.34}$ & 3.74$^{+0.40}_{-0.33}$& {5.14}$^{+0.52}_{-0.44}$& 3.78$^{+0.36}_{-0.15}$\\
\\[-0.95em]
&{\it Z} ({\it Z$_{\odot}$})& 0.23$^{+0.23}_{-0.16}$ & 0.23 (fixed)& 0.23 (fixed)& {0.14}$\pm{0.13}$&0.23 (fixed)\\
\\[-0.95em]
&{\it norm} (cm$^{-5}$)&2.99$^{+0.17}_{-0.22}$ $\times$ 10$^{-5}$ & 1.49$^{+0.20}_{-0.18}$ $\times$ 10$^{-3}$&1.52$^{+0.20}_{-0.18}$ $\times$ 10$^{-3}$&{3.00}$^{+0.16}_{-0.15}$ $\times$ 10$^{-3}$ &2.97$^{+0.09}_{-0.14}$ $\times$ 10$^{-5}$\\
\\[-0.95em]
&{\it (C or $\chi{^2}$)/ $\nu$} & 375.85 / 351 & 649.30 / 587& 651.16 / 587& 387 / 413&1695.97  / 1974\\
\\[-0.95em]
\hline
\\[-0.95em]
{B} & {\it kT} (keV) & 5.17$^{+3.51}_{-1.92}$& 3.55$^{+0.66}_{-0.47}$& 3.70$^{+0.61}_{-0.54}$& 4.72$^{+0.69}_{-0.59}$&3.55$^{+0.60}_{-0.45}$\\
\\[-0.95em]
&{\it Z} ({\it Z$_{\odot}$})& 0.18$^{+0.96}_{-0.18}$ & 0.18 (fixed)& 0.18 (fixed)& 0.09$^{+0.20}_{-0.09}$&0.18 (fixed)\\
\\[-0.95em]
&{\it norm} (cm$^{-5}$)&2.23$^{+0.47}_{-0.52}$ $\times$ 10$^{-5}$ & 1.51$^{+0.33}_{-0.29}$ $\times$ 10$^{-3}$& 1.43$^{+0.34}_{-0.24}$ $\times$ 10$^{-3}$& {1.91}$^{+0.17}_{-0.16}$ $\times$ 10$^{-3}$ &2.15$^{+0.28}_{-0.27}$ $\times$ 10$^{-5}$\\
\\[-0.95em]
&{\it (C or $\chi{^2}$)/ $\nu$} & 128.30 / 108&634.71 / 595 & 636.99 / 595& 211 / 252 & 1049.24 / 1133 \\
\\[-0.95em]
\hline
\\[-0.95em]
{C} & {\it kT} (keV) & 4.41$^{+1.02}_{-0.84}$ &3.00$^{+0.59}_{-0.41}$ & 3.17$^{+0.57}_{-0.41}$& {4.65}$^{+0.43}_{-0.41}$ &3.34$^{+0.44}_{-0.38}$\\
\\[-0.95em]
&{\it Z} ({\it Z$_{\odot}$})&0.44$^{+0.44}_{-0.27}$ &0.44 (fixed) & 0.44 (fixed)&  {0.17}$\pm{0.13}$&0.44 (fixed)\\
\\[-0.95em]
&{\it norm} (cm$^{-5}$)& 2.43$^{+0.30}_{-0.32}$ $\times$ 10$^{-5}$& 2.06$^{+0.50}_{-0.43}$ $\times$ 10$^{-3}$& 2.13$^{+0.46}_{-0.40}$ $\times$ 10$^{-3}$& {2.84}$^{+0.16}_{-0.15}$ $\times$ 10$^{-3}$ &2.37$^{+0.15}_{-0.16}$ $\times$ 10$^{-5}$\\
\\[-0.95em]
&{\it (C or $\chi{^2}$)/ $\nu$} & 299.07 / 253 &618.81 / 584 & 629.15 / 584&388 / 420 &1442.07 / 1652\\
\\[-0.95em]
\hline
\\[-0.95em]
{D} & {\it kT} (keV) &  4.95$^{+0.39}_{-0.38}$&4.01$^{+0.44}_{-0.37}$& 4.41$^{+0.39}_{-0.34}$&  {5.30}$^{+0.42}_{-0.33}$ &4.33$^{+0.28}_{-0.26}$\\
\\[-0.95em]
&{\it Z} ({\it Z$_{\odot}$})&0.25$^{+0.10}_{-0.11}$ &0.25 (fixed) &0.25 (fixed) & {0.21}$\pm{0.11}$ &0.25 (fixed)\\
\\[-0.95em]
&{\it norm} (cm$^{-5}$)& 4.89$^{+0.15}_{-0.19}$ $\times$ 10$^{-5}$&  2.23$^{+0.29}_{-0.26}$ $\times$ 10$^{-3}$& 2.56$^{+0.25}_{-0.23}$ $\times$ 10$^{-3}$& {4.60}$^{+0.19}_{-0.18}$ $\times$ 10$^{-3}$&5.03$^{+0.14}_{-0.13}$ $\times$ 10$^{-5}$\\
\\[-0.95em]
&{\it (C or $\chi{^2}$)/ $\nu$} & 685.84 / 611 &588.59 / 585& 615.74 / 585& 572 / 559& 2224.15 / 2548\\
\\[-0.95em]
\hline
\\[-0.95em]
{E} & {\it kT} (keV) & 5.16$^{+0.41}_{-0.37}$& 5.31$^{+0.53}_{-0.44}$&5.31$^{+0.49}_{-0.42}$  & {5.57}$^{+0.52}_{-0.40}$&5.00$^{+0.30}_{-0.28}$\\
\\[-0.95em]
&{\it Z} ({\it Z$_{\odot}$})& 0.37$^{+0.13}_{-0.10}$& 0.37 (fixed)& 0.37 (fixed)& {0.18}$\pm{0.12}$ &0.37 (fixed)\\
\\[-0.95em]
&{\it norm} (cm$^{-5}$)& 5.32$^{+0.22}_{-0.23}$ $\times$ 10$^{-5}$& 1.58$^{+0.14}_{-0.13}$ $\times$ 10$^{-3}$&1.60$^{+0.13}_{-0.12}$ $\times$ 10$^{-3}$ & {5.14}$\pm{0.22}$ $\times$ 10$^{-3}$&5.41$\pm{0.12}$ $\times$ 10$^{-5}$\\
\\[-0.95em]
&{\it (C or $\chi{^2}$)/ $\nu$} & 667.32 / 610 &659.01 / 580& 670.20 / 580& 498 / 47&2262.75 / 2502\\
\\[-0.95em]
\hline
\\[-0.95em]
{F} & {\it kT} (keV) & 4.81$^{+0.28}_{-0.24}$ & 5.27$^{+0.39}_{-0.34}$& 5.49$\pm{0.35}$& {5.65}$^{+0.54}_{-0.44}$&4.82$^{+0.22}_{-0.20}$\\
\\[-0.95em]
&{\it Z}({\it Z$_{\odot}$}) & 0.14$^{+0.06}_{-0.07}$ & 0.14 (fixed)& 0.14 (fixed)& {0.12}$\pm{0.12}$ &0.14 (fixed)\\
\\[-0.95em]
&{\it norm} (cm$^{-5}$)& 8.60$^{+0.18}_{-0.21}$ $\times$ 10$^{-5}$& 2.62$^{+0.19}_{-0.18}$ $\times$ 10$^{-3}$& 2.79$^{+0.19}_{-0.16}$ $\times$ 10$^{-3}$ & {7.00}$^{+0.30}_{-0.29}$ $\times$ 10$^{-3}$&8.73$^{+0.11}_{-0.14}$ $\times$ 10$^{-5}$\\
\\[-0.95em]
&{\it (C or $\chi{^2}$)/ $\nu$} & 919.68~/~890 & 621.19 / 594 & 630.87 / 594 & 433 / 450 & 2631.39 / 2926 \\
\enddata
\tablenotetext{a}{$\chi{^2}$ is used.}
\tablenotetext{b}{Scattered light treatment is included.}
\end{deluxetable*}

An additional ${\tt apec}$ component improved the statistics of the Region E spectrum with respect to the single ${\tt apec}$ model by {\it $\Delta$C/$\Delta\nu$} = 19.42/2. This two-temperature model resulted in a high-temperature component of {\it kT} = 15.91$^{+15.92}_{-4.70}$ keV and a low-temperature component of {\it kT} = 3.78$^{+0.37}_{-0.32}$ keV. Instead, when the ${\tt apec}$ + ${\tt powerlaw}$ model was used, the statistics were improved by {\it $\Delta$C/$\Delta\nu$} = 24.41/2. This combined model resulted in a power-law emission with a photon index of $\Gamma$ = 1.80$^{+0.12}_{-0.17}$, and the thermal component was {\it kT} = 4.28$^{+0.57}_{-0.27}$ keV.

Since we did not find a constrained, strong high-temperature component in Region A, we restricted our spectral analysis to a smaller region with r = 1$\arcmin$.5 centered at the X shown in Fig.~\ref{fig:6reg}.

We fitted an ${\tt apec}$ model to the joint \nustar\ and \xmm\ spectra and the \xmm\ spectrum alone. For these analyses, we used both fixed and free abundance and {\it N$_H$} parameters, having eight different spectral fit. The abundance was fixed to {\it Z} = 0.23 {\it Z$_{\odot}$} as obtained from the \xmm\ spectral fit for Region A. We present the results of this analysis in Section~\ref{sec:hotspotfit}.

\begin{figure}
\centering
\includegraphics[width=86mm]{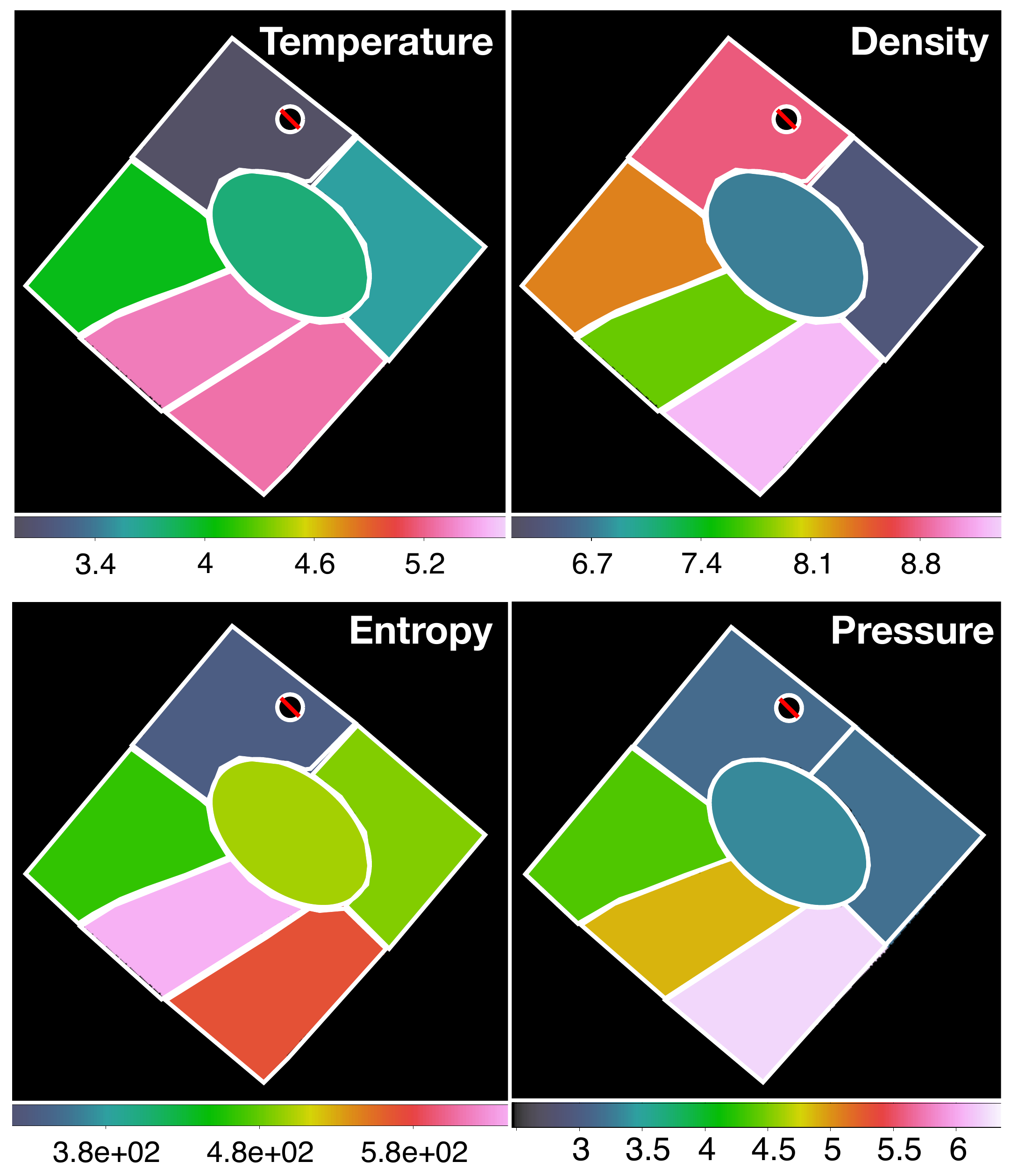}
\caption{\nustar\ temperature ({\it upper left}), deprojected density ({\it upper right}), deprojected entropy ({\it lower left}) and deprojected pressure maps ({\it lower right}) for regions of interest. Temperature units are given in keV, density is given in 10$^{-4}$ cm$^{-3}$ scaled by  (LOS/1 Mpc)$^{-1/2}$, entropy is in keV cm$^{2}$ scaled by  (LOS/1 Mpc)$^{1/3}$, and pressure is presented in 10$^{-12}$ dynes cm$^{-2}$ scaled by (LOS/1 Mpc)$^{-1/2}$. \label{fig:NuSTARprofiles}}
\end{figure}

We also created deprojected thermodynamical maps using the ${\tt XSPEC}$ model ${\tt apec}$, which is defined as $\left [10^{-14}/4\pi \left [D_{A}(1+z) \right ]^{2} \right ]\int n_{e}n_{H}dV$, to obtain electron density n$_{e}$, with the assumption of a fully ionized plasma, {\it n$_{e}$} $\simeq$ 1.2 {\it n$_{H}$}. For the volume, we used the area of the ROI  obtained using SAOImageDS9\footnote{\url{https://sites.google.com/cfa.harvard.edu/saoimageds9}} multiplied by the depth, and for the depth we assumed 1 Mpc for the line of sight (LOS), therefore the density values are scaled by (LOS/1 Mpc)$^{-1/2}$ similar to the approach adopted by \citet{akamatsu17}. For entropy and electron pressure, we used the commonly adopted definitions, {\it S} = {\it kT} $\times$ {\it n$^{-2/3}_{e}$} (scaled by (LOS/1 Mpc)$^{1/3}$) and {\it P$_{e}$} = {\it n$_{e}$} $\times$ {\it kT} ((LOS/1 Mpc)$^{-1/2}$), respectively \citep{gitti10}. Gas pressure then becomes {\it P} = {\it n} $\times$ {\it kT}, where we assume n = 2n$_{e}$. The resulting maps are shown in Fig.~\ref{fig:NuSTARprofiles}, and the corresponding errors are given in Table~\ref{tab:thermodynamic}.

We made two more assumptions for LOS for our thermodynamical values to be easily compared with future work. Firstly, we adopted r$_{180}$ = 34$\arcmin$.6 \citep{markevitch98} and r$_{180}$ = r$_{vir}$, and estimated the distance of the center of each region from the midpoint of subcluster centers of A3395. The distances (x) we find for regions A, B, C, D, E, and F are 637 kpc ($\sim$0.32 r$_{vir}$), 876 kpc ($\sim$0.44 r$_{vir}$), 776 kpc ($\sim$0.39 r$_{vir}$), 438 kpc ($\sim$0.22 r$_{vir}$), 338 kpc ($\sim$0.17 r$_{vir}$), and 458 kpc ($\sim$0.23 r$_{vir}$), respectively.

These estimated distances are used for a secondary estimation of LOS. Along the center of the cluster where x=0 and $\theta$=0, the depth is 2r$_{500}$ $\sim$2 Mpc, and then it scales with varying [x,$\theta$]. The given distances (x) for these regions can be used as {\it sin}($\theta$), where the LOS is the 2{\it cos}($\theta$) assuming a perfect sphere with r = r$_{vir}$. With these assumptions, LOS for regions A, B, C, D, E, and F becomes $\sim$1.33 Mpc, $\sim$0.62 Mpc, $\sim$1.02 Mpc, $\sim$1.64 Mpc, $\sim$1.73 Mpc, and $\sim$1.62 Mpc, respectively.

A third estimation for LOS is achieved by using r$_{vir}$ instead of r$_{500}$. This time, along the center of the cluster where x=0 and $\theta$=0, the depth is 2r$_{vir}$ $\sim$4 Mpc, and then it again scales with varying [x,$\theta$]. We now assume  r = r$_{vir}$. With these assumptions, LOS for regions A, B, C, D, E, and F becomes $\sim$3.76 Mpc, $\sim$3.56 Mpc, $\sim$3.68 Mpc, $\sim$3.90 Mpc, $\sim$3.92 Mpc, and $\sim$3.89 Mpc, respectively.

Using these physical distances along with the scaling factors used for density ((LOS/1 Mpc)$^{-1/2}$), entropy ((LOS/1 Mpc)$^{1/3}$) and pressure ((LOS/1 Mpc)$^{-1/2}$), results presented in Table~\ref{tab:thermodynamic} can easily be used with future work depending on the choice of radius for the cluster emission.

\begin{deluxetable*}{lcccccc}
\tabletypesize{\scriptsize}
\tablewidth{0pt} 
\tablecaption{Deprojected thermodynamical parameters from \nustar\ (3.0-15.0 keV) spectral fits for the regions of interest where the scattered light emission is included.
\label{tab:thermodynamic}}
\tablehead{\\[-0.95em]
& \colhead{Region A} & \colhead{Region B} & \colhead{Region C} & \colhead{Region D} & \colhead{Region E} & \colhead{Region F}}
\startdata\\[-0.95em]
{\it Area} (arcmin$^{2}$) & 25.4 & 28.5 & 20.8 & 24.9 & 20.3 & 23.9 \\
\\[-0.95em]
{\it n$_{e}$} (10$^{-4}$ cm$^{-3}$) & 6.74$^{+0.42}_{-0.41}$ & 6.40$^{+0.61}_{-0.70}$ & 8.75$^{+1.06}_{-0.91}$ & 8.32$^{+0.54}_{-0.49}$ & 7.77$^{+0.34}_{-0.32}$ & 9.21$^{+0.33}_{-0.32}$ \\
\\[-0.95em]
{\it S} (keV cm$^{2}$) & 486$^{+56}_{-48}$ & 478$^{+95}_{-70}$ & 328$^{+70}_{-50}$ & 453$^{+53}_{-45}$ & 628$^{+65}_{-55}$ & 557$^{+43}_{-038}$ \\
\\[-0.95em]
{\it P} (10$^{-12}$ dynes cm$^{-2}$) & 4.04$^{+0.51}_{-0.44}$ & 3.64$^{+0.78}_{-0.59}$ & 4.21$^{+0.97}_{-0.72}$ & 5.35$^{+0.68}_{-0.58}$ & 6.61$^{+0.72}_{-0.61}$ & 7.77$^{+0.64}_{-0.57}$\\
\\[-0.95em]
\enddata
\end{deluxetable*}

\begin{figure}
\centering
\includegraphics[width=83mm]{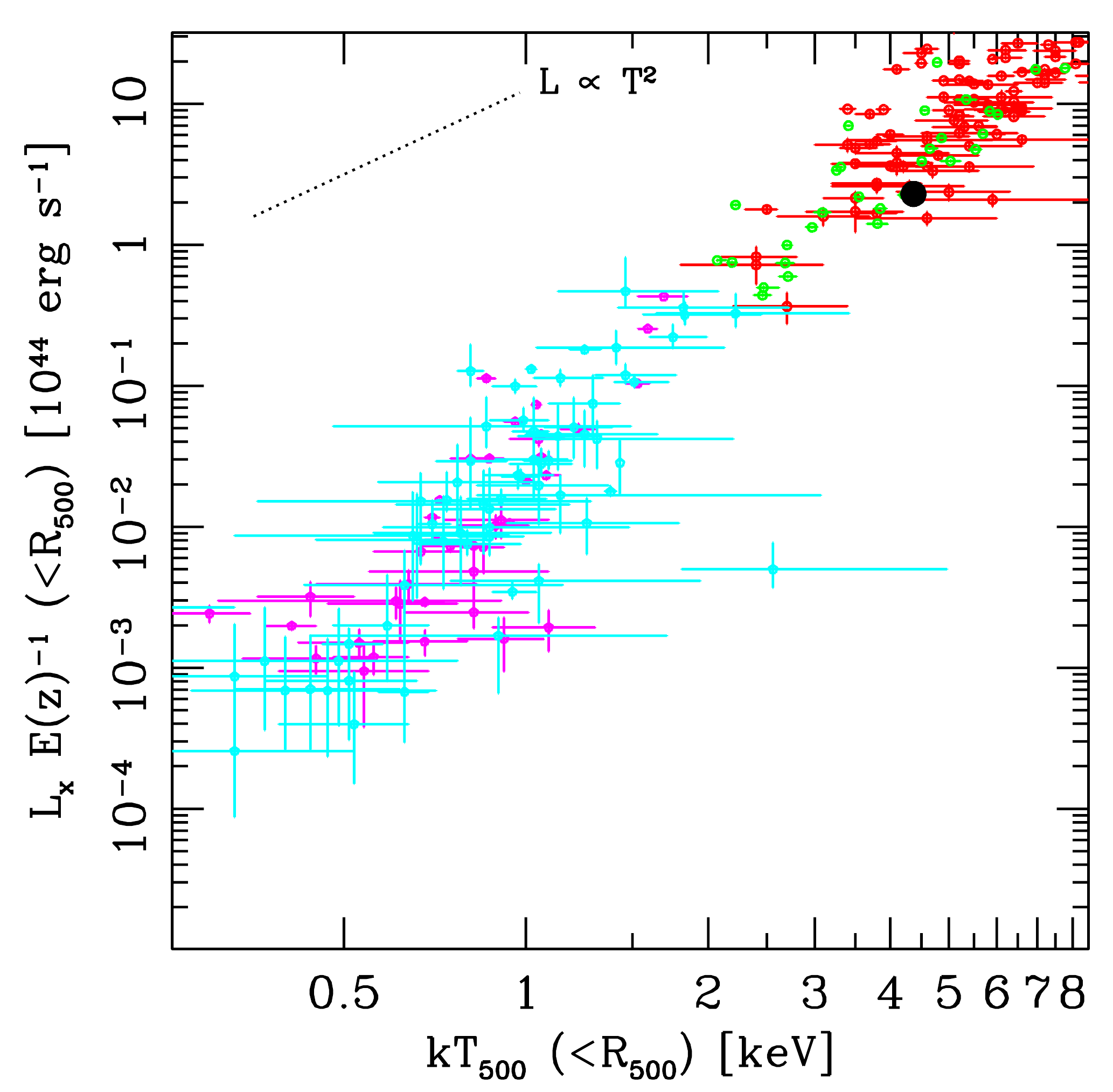}
\caption{X-ray bolometric luminosity vs. X-ray temperature (adapted from \citealt{gaspari14}: see this paper for the groups/clusters samples included with different colors). A3395 is superimposed with a black dot. $E(z)$ is the cosmological evolution factor, near unitary value.
It clearly touches the lower envelope of the scaling relation, which is the typical location of non-cool-core systems.  
\label{fig:Gaspari14}}
\end{figure}

We also calculated the luminosity of the cluster from the \xmm\ spectra within r = 796$\arcsec$ = 0.83 R$_{500}$ adopting R$_{500}$ = 930 kpc = 954$\arcsec$ \citep{alvarez18}. This selected region covers the whole FOV of the \xmm\ observation. We find the X-ray bolometric (0.01-100 keV) luminosity to be 
{\it L$_{X}$} = 2.342$^{+0.015}_{-0.011}$ 10$^{44}$ erg s$^{-1}$, and within 0.5-2.0 keV we find {\it L$_{X}$} = 1.204$^{+0.041}_{-0.071}$ 10$^{44}$ erg s$^{-1}$
using ${\tt XSPEC}$ convolution model ${\tt clumin}$. Within this region, we find {\it kT} = 4.37$\pm{0.04}$ keV for the 0.5-7.0 keV band using a single-temperature model. In Fig.~\ref{fig:Gaspari14}, we added the core X-ray properties of A3395 (black dot) to the bolometric luminosity vs. X-ray temperature scaling relation presented in \citet{gaspari14}.
We note that since the error bars are small, they are contained within the black dot, which is enlarged to facilitate the differentiation from the rest of the sample.

\subsection{The point source in Region C}

We also analyzed the spectral properties of the point source residing inside Region C previously cataloged as 2MASX J06261214-5417071 at z = 0.05050 (SIMBAD, \citet{donnely01,paturel03,jones09,marziani17}). We extracted a circular region of r = 1$\arcmin$.2 centered on the point source. We find that the best-fit model that describes the emission from the source in the 3-20 keV band is an ${\tt apec}$ + ${\tt powerlaw}$ model with a photon index of $\Gamma$ = 1.48$^{+0.56}_{-0.83}$, and {\it C/$\nu$} = 412.10/457. The plasma temperature was found to be {\it kT} = 1.81$^{+0.95}_{-0.59}$ keV, with {\it Z} = 0.43$^{+2.47}_{-0.31}$ {\it Z$_{\odot}$}. The luminosity of the ${\tt powerlaw}$ component is found to be {\it L$_{X}$} = 1.434$^{+0.153}_{-0.166}$ 10$^{42}$ erg s$^{-1}$ within 3.0-20.0 keV and {\it L$_{X}$} = 5.962$^{+0.706}_{-0.676}$ 10$^{40}$ erg s$^{-1}$ within 0.5-2.0 keV. For the ${\tt apec}$ component, we found 3.0-20.0 keV luminosity to be {\it L$_{X}$} = 7.330$^{+1.008}_{-0.977}$ 10$^{41}$ erg s$^{-1}$ and the 0.5-2.0 keV luminosity is {\it L$_{X}$} = 3.333$^{+0.458}_{-0.444}$ 10$^{41}$ erg s$^{-1}$.

This two-component model improved the statistics by with respect to a single-temperature model by {\it $\Delta$C/$\Delta\nu$} = 14.15/2. The single-temperature model resulted in {\it kT} = 6.44$^{+0.86}_{-0.74}$ keV plasma with an unconstrained abundance value. When we freed the redshift, we found {\it z} = 0.061$^{+0.025}_{-0.028}$, which agrees with both the cluster and source (2MASX J06261214-5417071, \citet{jones09}) redshifts  within 1$\sigma$.

We extracted a spectrum using the same region from \xmm\ data. We grouped the \xmm\ spectrum by 3 as well and applied ${\tt cstat}$ for direct comparison with \nustar\ results. A single-temperature model indicates {\it kT} = 4.64$^{+0.60}_{-0.37}$ keV plasma with {\it Z} = 0.58$^{+0.25}_{-0.30}$ {\it Z$_{\odot}$}. An addition of ${\tt powerlaw}$ component did not improve the fit and the photon index was not constrained. However, a single ${\tt powerlaw}$ model fit without an ${\tt apec}$ component gives a photon index of $\Gamma$ = 1.90$^{+0.05}_{-0.09}$. Since these two emission models can be degenerate within \xmm\ operating energy band given the similar statistics, we extracted a spectrum in the vicinity of the point source with the same area, and characterized the ICM within that region. A single-temperature model in this region showed a plasma with {\it kT} = 3.04$^{+1.83}_{-0.76}$ keV  and {\it Z} = 0.36$^{+0.87}_{-0.25}$ {\it Z$_{\odot}$}. The parameters obtained from the ${\tt apec}$ fit of the region in the vicinity of the point source were then inserted and fixed during the fitting procedure of the point source, where a combined ${\tt apec}$ + ${\tt powerlaw}$ model was implemented. We then found a photon index of $\Gamma$ = 1.86$^{+0.10}_{-0.11}$ for the ${\tt powerlaw}$ component.  Within 0.5-2.0 keV, we estimate a luminosity of {\it L$_{X}$} = 3.555$^{+0.308}_{-0.279}$ 10$^{41}$ erg s$^{-1}$ again for the ${\tt powerlaw}$ component.

\section{Discussion} \label{sec:discussion}

We studied \nustar, \xmm, and \suzaku\ data of the merging cluster A3395. In this section, we discuss our results in relation to the literature.

\subsection{On the global properties}
For the \nustar\ data of A3395, we first considered
the global spectrum of the part of the cluster within
the FOV, which was fit to a single temperature thermal
plasma model ({\tt apec}).
We found an average temperature of $kT = 5.59 \pm 0.11$~keV. Although this is the global result form \nustar\ FOV, the pointing only covers the northwestern part of the cluster, and therefore this is not the global temperature of the cluster itself. However, the lack of a good fit suggests a single-temperature model insufficiently describes the data. As this is a merging cluster, we cannot assume isothermality across the FOV, and we expect the FOV to be contaminated by scattered light due to the presence of the subcluster centers in the vicinity of the pointing. We therefore fit this spectrum with two more models, with either an additional ${\tt apec}$ or ${\tt powerlaw}$ component.

We find that the ${\tt apec}$+${\tt powerlaw}$ model better describes the global spectrum with {\it kT} = 2.06$^{+0.31}_{-0.22}$ keV and {\it $\Gamma$} = 1.82$^{+0.18}_{-0.29}$.
However, this $\simeq$ 2 keV temperature is much smaller than expected based on the results of \citet{markevitch98}, \citet{donnely01} and \citet{lakhchaura11}. In addition, the single-temperature model result of $\simeq$ 5.6 keV, is then higher than the global temperatures found for the cluster in the literature as found by \citet{markevitch98,donnely01,lakhchaura11,alvarez18}. The global temperatures they report account for the hot plasma at the cluster center and the \nustar~FOV does not cover the central region of A3395. The cluster temperatures are expected to drop with respect to increasing radius assuming there are no shocks. The $\simeq$ 5.6 keV result we find for \nustar~FOV should be even lower than their reported values and not higher. Although \citet{lakhchaura11} points to high-temperature regions enclosed in our \nustar\ FOV, which gave a motivation to study these regions in detail with \nustar\, we recall that they also state a 60\% error on these values. 

Further investigating the impact of scattered light, we found
that its contribution could also be modeled with a {\tt powerlaw}
model. This fact, combined with the presence of multi-temperature gas,
suggests that the ${\tt apec}$+${\tt powerlaw}$ model is sufficiently
flexible to capture these more extensive components and that
the ${\tt powerlaw}$ component should not be interpreted to have
physical meaning. Therefore, taking advantage of \nustar's imaging capability, we continued with a grid analysis as described in Section~\ref{sec:grid}. This analysis and the \xmm\ photon image (upper panel of Fig.\ref{fig:6reg}) showed that there are six regions in the \nustar\ FOV with similar thermodynamical properties and substructures.

We also studied the \xmm\ FOV that covers the central $\sim$0.83 R$_{500}$ of the cluster to obtain the global temperature and luminosity of the cluster. Our temperature result {\it kT} = 4.37$\pm{0.04}$ keV, which is in agreement with the literature \citep{markevitch98,donnely01} within 1$\sigma$ errors. We found a luminosity of {\it L$_{X}$} = 1.204$^{+0.041}_{-0.071}$ 10$^{44}$ erg s$^{-1}$ within 0.5-2.0 keV, which is in agreement with the luminosity estimation of \citet{degrandi99} within 1$\sigma$. 
We found the X-ray bolometric (0.01-100 keV) luminosity to be 
{\it L$_{X}$} = 2.342$^{+0.015}_{-0.011}$ 10$^{44}$ erg s$^{-1}$, as plotted in the {\it L$_{X}$}-{\it T$_{X}$} scaling relation in Fig.~\ref{fig:Gaspari14}. Cool-core systems tend to reside in the upper envelope of the scaling relation due to a higher {\it L$_{X}$} per given {\it T$_{X}$} (or {\it M$_{\rm total}$} since {\it T$_{X}$} is a tight proxy for cluster mass), whereas merging clusters whose cool cores have been disrupted reside in the lower envelope \citep{gaspari14}. A3395 lies at the lower boundary, in agreement with the typical behavior of the population of non-cool core clusters, which do not have the inner cool region of the ICM. This suggests that A3395 might have evacuated significant gas mass (moving toward the bottom), heated the gas (moving toward the right), or this cluster has assembled in a poor gas environment. Having a hot ($\sim$6 keV) intracluster filament in between its subclusters (Region E in our analysis, Region 2 in \citet{markevitch98}, Region 3 in \citet{donnely01}, Region F in \citet{lakhchaura11}), and lacking a cool core, A3395 appears to be still in an early stage of merger, since it lacks a major overheating. 

\subsection{On the treatment of scattered light}
Using ray-trace simulations with MT\_RAYOR, we assessed the scattered light contamination in our observation. We find that the scattered light is best modeled with a ${\tt powerlaw}$, and we provide a quantitative description of this contamination. This is an empirical model and due to the nonuniformity of the scattered light contamination, its complete behavior needs to be assessed with further studies of the off-axis angle and source energy dependence, as well as the position on the detector. This is a multidimensional problem, and our method is the only known approach to study \nustar\ data that has scattered light contamination, to the best of our knowledge. The method is described in detail in Section~\ref{sec:SL}, and the method is summarized in flow chart shown in Fig.~\ref{fig:recipe}.

We find that in Regions A and B, where A3395 connects with an intercluster filament, the scattered light contamination is below 5\% of the total flux. Regions C, D, and F suffer from scattered light at the $\sim$15\% level, and this contamination rises to $\sim$25\% for Region E. This is expected since Region E is near both of the bright subcluster centers, namely the NE and SW regions denoted in the upper panel of Fig.~\ref{fig:6reg}.

A quick comparison of $C$-stat values of \nustar~ spectral fits with and without SL treatment in Table~\ref{tab:allfit} shows that the fits improved for all ROIs when the scattered light component is included, with $\Delta C$ ranging from $\sim$2 
to $\sim$27. Thus, the inclusion of our scattered light as an additional background component results in a better assessment of the source emission. However, we note that for all ROIs, the \nustar\ temperature results from the spectra with and without scattered light treatment agree within 1$\sigma$. Our temperature results from the \nustar\ data are in agreement to those from \xmm\, and joint \nustar\ and \xmm\ analyses for a region with $\sim$25\% scattered light contamination within 1$\sigma$. Our results show that, for regions of interest where scattered light contamination is above 10\% (Regions C, D, E, and F), the temperature values are in agreement within 1.6$\sigma$ for \nustar, \xmm, \suzaku, and joint \xmm\ and \nustar\ spectral fits, which validates our approach to tackling the \nustar\ scattered light contamination for this observation.

We claim that, although the effect of scattered light contamination depends on the flux and emission features of structures, at a level of up to $\sim$25\%, we seem to be safe within 1$\sigma$ errors of the face value of temperature. However, we also note that further investigation is needed to fully understand the effect of scattered light at various plasma temperatures. Our technique can be used for future \nustar\ observation proposals to estimate the possible scattered light contamination.

\subsection{On the regions of interest}

\nustar~temperature results from all regions of interest seem to be lower than what is found with \suzaku~(Table~\ref{tab:allfit}). This trend is also seen in regions A, B, C, and D for \nustar~versus \xmm. Cross-calibration studies similar to \xmm~and \chandra~by \citet{schellenberger15} are required for \nustar, \xmm, and \suzaku~to understand this behaviour.

In our detailed analysis of the possible connection region of the cluster with the intercluster filament, Region A, we cannot confirm the existence of a strong high-temperature component in any of our \nustar, \xmm, \suzaku, and joint \nustar\ and \xmm\ fits, as previously reported in the temperature map of \citet{lakhchaura11}. We consistently find temperature values around 4--5~keV in all datasets considered. This region is also present in a temperature map based on {\it ASCA} observations \citep{markevitch98}, and our temperature results from \nustar, \xmm, \suzaku, and joint \nustar\ and \xmm\ fits all agree with their results within 1$\sigma$. However, \citet{lakhchaura11} do note 60\% errors on their map for regions lying at the edge of the FOV of the \xmm\ pointing due to the low S/N.

Although we find a high-temperature component ($\sim$16 keV) for Region A with \nustar\ analysis, where two-temperature plasma model was applied to the spectrum, only a lower bound ($\sim$8 keV) for this higher-temperature component is found; the ICM is mainly dominated by the cooler component ($\sim$4 keV). 

To better isolate a possible hot spot lying in Region A, we extracted spectra in a smaller region ($r = 1.5\arcmin$) from both \nustar, and \xmm, centered on the hottest region in the \xmm\ temperature map reported by \citet{lakhchaura11}. During these fitting procedures, we investigated the effect that column density and abundance may have on the temperature measurements, due to the possible bias against \xmm~ temperatures caused by the uncertainties in the effective area at soft energy \citep{schellenberger15}. While keeping the \textit{N$_{H}$} frozen, we first kept the \textit{Z} parameter fixed to the value found from the \xmm~analysis of Region A. Then we fit the spectra again by allowing \textit{Z} parameter to be free. We repeated the same process by freeing the \textit{N$_{H}$}.

The highest temperature we find is $kT = 5.13^{+1.80}_{-1.48}$~keV with our \xmm\ analysis (Table~\ref{tab:centerXfitxmm}). We found lower temperatures with a joint \xmm~and \nustar~analysis than with the \xmm\ analysis alone (Table~\ref{tab:centerXfitxmmnustar}). In addition, all temperature values from these procedures agree within 1$\sigma$. Since background dominates at the edge of the {\bf \xmm~}FOV, we hypothesize that the discrepancy between our results and the literature at the hot spot location may be due to how the background was treated; for example, the background model of \citet{lakhchaura11} might underestimate the true background at those locations.

Region B is also located near the connection region of A3395 and the intercluster filament. This region was also partially covered by the analysis of \citet{markevitch98}, whose temperature results agree with ours at the 1$\sigma$ level. Regions C and D are at least partly included in the NW region studied by \citet{lakhchaura11}, where our \xmm\ temperature result agrees with their findings within 1$\sigma$.

The bridge emission between the two subclusters of A3395 enclosed by our Region E shows high temperature, entropy, and pressure with respect to the surrounding ICM. Moreover, this region was well fit by a two-temperature plasma model. This bridge is thought to be ram pressure-stripped from the northern subcluster in the A3395 merging system. \citet{lakhchaura11} finds a higher temperature for this {\it intracluster} filament than we do, yet our temperature value is within 1.5$\sigma$ of their value.

In addition, we searched for a nonthermal X-ray counterpart of the faint extended radio source to the west of A3395, lying in our Region F \citep{erosita21}. They argue they this source denoted as S2/S3 in their work, may be a radio relic or may due to reaccelerated relativistic plasma. We find no significant nonthermal emission in this region, possibly because the hot ICM dominates the emission.

The electron entropy is closely related to the thermodynamical history of the clusters \citep{voit05}. In particular, the entropy of the ICM decreases in the process of radiative cooling and increases when heating energy is introduced into the ICM, e.g., via merging and feedback processes (\citealt{gaspari15}). And at the interface regions of cluster outskirts and WHIM filaments is the zone where the entropy flattening is observed \citep{alvarez18}. To assess the merger history of the cluster and possible interaction of the filament and the cluster, we estimated the entropy for our ROIs. Since it is difficult to create radial profiles of a sample of non-cool-core systems due to asymmetrical morphology as well as nonthermal processes caused by mergers, we compared the entropy values with the 13 nearby cooling-flow cluster entropy profiles studied by \citet[][Figure 5]{piffaretti05} in the following paragraph. 

In order make this comparison, we used the distance estimations of regions explained in Section~\ref{sec:allfit}.
For Region A at $\sim$0.32 r$_{vir}$, we find the entropy to lie below the fitting curve yet within the scatter of the sample, and in addition, within 1$\sigma$ of galaxy cluster 2A 0335+096. Region B, at $\sim$0.44 r$_{vir}$, again is at the lower boundary of the scatter, in agreement with the entropy of Sersic 159-3 at the same distance from the core. Region C (at $\sim$0.39 r$_{vir}$), has a lower entropy than the whole sample range. The entropy of Region D at $\sim$0.22 r$_{vir}$ is within the sample entropy values, still lying below the mean. The entropy of Region E ($\sim$0.17 r$_{vir}$), lies above the fitted curve. Finally, the entropy value of Region F ($\sim$0.23 r$_{vir}$), seems to lie above the fitted curve as well. All regions except for Region C have similar entropy with cooling-flow cluster entropy profiles, where Regions A, B, and D lie below the fitted curve, and Regions E and F lie above.

The high entropy and the high temperature of Region E (the bridge) indicate a heating process that may be caused by the gravitational pull of the ICM from subclusters that are at a pre-merger stage.

In addition to X-ray studies, the ICM of A3395 has been studied through the SZ effect with \planck. \citet{planck13} report temperature and pressure values for the subclusters in A3395. Our Region D is enclosed in their A3395E region, and Region F is enclosed in Planck A3395SW region. Their GNFW2 pressure profile model results in {\it kT} = 5.0 keV and {\it P} = 0.40 $\times$ 10$^{-2}$ keV/cm$^{3}$ for A3395E (Region D), and {\it kT} = 4.8 keV and {\it P} = 0.40 $\times$ 10$^{-2}$ keV/cm$^{3}$ for A3395SW (Region F). With our \nustar\ analysis we find 4.01$^{+0.44}_{-0.37}$ and {\it P} = 0.33$\pm{0.04}$  $\times$ 10$^{-2}$ keV/cm$^{3}$ (5.35$^{+0.68}_{-0.58}$ $\times$ 10$^{-12}$ dynes cm$^{-2}$) for Region D, and 5.27$^{+0.39}_{-0.34}$ keV and {\it P} = 0.48$^{+0.04}_{-0.03}$ $\times$ 10$^{-2}$ keV/cm$^{3}$ (7.77$^{+0.64}_{-0.57}$ $\times$ 10$^{-12}$ dynes cm$^{-2}$) for Region F. They do not report uncertainties for these specific regions (being model-dependent on global fits).       

Although it is difficult to make a direct comparison between \nustar\ and \planck\ analyses since the derivation of pressure is based on the different methods, it is important to state that our result reaches the \planck\ value within 1$\sigma$ errors. This is also a good validation of our scattered light treatment, as well as the assumptions used for the deprojection of thermodynamical maps, since the estimation of these parameters is the end product of multiple treatments, assumptions, and analyses.

We visually detected a point source in both \nustar\ and \xmm\ images, and extracted a circular region with r = 1$\arcmin$.2 region to study the source in detail from both \nustar\ and \xmm\ data. We found that a ${\tt powerlaw}$ component with $\Gamma$ = 1.48$^{+0.56}_{-0.83}$ and a thermal plasma with {\it kT} = 1.81$^{+0.95}_{-0.59}$ keV using \nustar\ data best describe the emission. For \xmm\ analysis, we selected a region with the same area in the vicinity of this source due to the degeneracy of the ${\tt powerlaw}$ and ${\tt apec}$ components within \xmm\ bandpass. This analysis helped in estimating the plasma properties in the vicinity, and resulting parameters were adopted and fixed in the fit of the point source region, which resulted in a ${\tt powerlaw}$ component with $\Gamma$ = 1.86$^{+0.10}_{-0.11}$, which is better constrained than the \nustar~ photon index, since all apec parameters were frozen during the \xmm~fit. We also tried using an additional ${\tt apec}$ model instead of ${\tt powerlaw}$, yet then the secondary ${\tt apec}$ temperature was around 5 keV, and the statistics were comparable, i.e., {\it C/$\nu$} = 570.35/702 for ${\tt apec}$ and {\it C/$\nu$} = 569.03/702 for ${\tt powerlaw}$. This $\sim$5 keV value is higher than what we find in Region C, where the point source resides, with \nustar, \xmm, and \suzaku, as well as joint \nustar\ and \xmm\ results, pointing to the degeneracy of a {\it kT} $\sim$ 5 keV thermal emission and $\Gamma$ $\sim$1.86 power-law emission for the \xmm\ analysis. Since \nustar\ covers a wider energy range, we claim that the true emission includes a power-law emission as well as a thermal emission within that region. The photon indices obtained from the two analyses agree within 1$\sigma$, therefore we cannot claim a statistically significant variability between \nustar\ and \xmm\ observations. When the \xmm~and \nustar~spectra were simultaneously fit using a {\tt powerlaw} + {\tt apec} model, the {\tt apec} component dominated the spectra, as expected due to the inclusion of more lower energy \xmm~photons in the spectra.

In addition, results from \nustar\ spectral analysis for the temperature and the luminosity of the gas confined within this r = 1$\arcmin$.2 suggest that the point source may be a thermal corona embedded in a hot environment \citep{sun07,tumer19}. The photon indeces of $\Gamma$ = 1.86$^{+0.10}_{-0.11}$ for \xmm\ and $\Gamma$ = 1.48$^{+0.56}_{-0.83}$ favor an AGN emission rather than X-ray binaries ($\Gamma$ $\leq$ 1.4) \citep[see, e.g.,][]{tozzi06}. Furthermore, ${\tt powerlaw}$ component accounts for the $\sim$66\% of the total luminosity within 3.0-20.0 keV, and $\sim$15\% within 0.5-2.0 keV based on \nustar\ analysis.

\subsection{On the intercluster filament}
Guided by the literature \citep{tittley01,lakhchaura11,planck13,bourdin20,erosita21}, we studied Regions A, B, and C in detail with the assumption that these regions may represent an interface of the A3395 ICM and the intercluster filament.
\citet{alvarez18} find a global temperature of $kT = 4.45^{+0.89}_{-0.55}$~keV and density $n_{e} = 1.08^{+0.06}_{-0.05} \times 10^{-4}$~cm$^{-3}$ for the intercluster filament. The temperature results obtained from \nustar, \xmm, \suzaku, and the joint \nustar\ and \xmm\ spectral analysis of Regions A and B, are in agreement with \citet{alvarez18} for the filament within 1$\sigma$. However, for Region C, the temperature results from \nustar\ analysis show a cooler plasma than what is found for the filament, yet in agreement within 1.3$\sigma$ of \citet{alvarez18}, and within 1$\sigma$ with our \xmm, \suzaku, and the joint \nustar\ and \xmm\ analyses. In addition, the density of the filament found by \citet{sugawara17} and \citet{alvarez18} is much smaller ($\sim1/4$) than the density we find for Regions A, B, and C. 

The entropy is expected to rise to values higher than 1000 keV cm$^{2}$ outside 0.5 R$_{200}$ \citep{pratt06}, due to heating by accretion shocks, and \citet{lakhchaura11} report high entropy and high temperature values for these regions. However, we do not observe such high entropy or temperature in Regions A, B, and C, which are expected to have higher entropy than the regions close to the center of the cluster \citep{piffaretti05}. In addition, the entropy we find for these regions are even lower than what is found at that similar radii for cool-core clusters \citet{piffaretti05}. 

These results, when studied in conjunction with the low temperature, low pressure, and high density values, suggest an excess of radiative cooling, which points to a flow of ICM into the filament, in contrast to \citet{erosita21}, who find high-temperature gas in the interface region and suggest heating by shocks via the ongoing merger activity.
Such an offset cooling process is analogous to the more vigorous multiphase condensation `weather' occurring in the dense cluster cores (\citealt{gaspari18}). Indeed, mergers drive significant amount of turbulent motion, which can locally enhance density (\citealt{gaspari13}) and thus lead to localized enhanced filamentary cooling (\citealt{wittor20}).
Such detections of cooling in merger systems have become more frequent in recent years (e.g., \citealt{somboonpanyakul21}).
Our results are also in line with \citet{alvarez18}, 
who suggest that the ICM gas in the outskirts may be tidally moved into the filament during the interaction, as a part of the merging processes of A3395 and A3391. Such tidal motions can be seen as another form of large-scale turbulence, with the related eddies locally enhancing density.

\section{Conclusion} \label{sec:conclusion}
We observed A3395 with \nustar\ for a total exposure time of $\sim$125 ks. We studied the northwestern region of the cluster in junction with archival \suzaku\ and \xmm\ observations.
We find that the location of the cluster that meets the intercluster filament does not show any signs of heated plasma, on the opposite, but rather shows signs of excessive cooling. This is likely linked to the condensation `weather' enhanced by turbulence or tidal motions, in analogy to the core counterparts (e.g., \citealt{gaspari20}).

In addition, our temperature results from the \nustar\ data are in agreement with those from \xmm\ and joint \nustar\ and \xmm\ analysis for a region with $\sim$25\% scattered light contamination within 1$\sigma$, and we claim that temperature assessment of the intracluster medium is still valid even when the data are contaminated up to $\sim$25\%. Our technique can be used for future \nustar\ observation proposals to estimate the possible scattered light contamination, and for its quantification during ICM analysis from moderately contaminated \nustar\ data.
\\

We thank the anonymous referee for valuable discussions and suggestions that improved our work significantly. This research has made use of data from the \nustar\ mission, a project led by the California Institute of Technology, managed by the Jet Propulsion Laboratory (JPL), and funded by by the National Aeronautics and Space Administration (NASA); \xmm\, an ESA science mission with instruments and contributions directly funded by ESA Member States and the USA (NASA); \suzaku\ satellite, a collaborative mission between the space agencies of Japan (JAXA) and the USA (NASA). In this work, we used the NuSTAR Data Analysis Software (NuSTARDAS) jointly developed by the ASI Science Data Center (ASDC, Italy) and the California Institute of Technology (USA). The data for this research have been obtained from the High Energy Astrophysics Science Archive Research Center (HEASARC), provided by NASA’s Goddard Space Flight Center. A.T. thanks Fiona A. Harrison, Kristin K. Madsen, and Brian W. Grefenstette for valuable discussions on scattered light, and Herv\'e Bourdin and Randall A. Rojas Bolivar for their contribution on the \nustar\ proposal process. A.T. and D.R.W. acknowledges support from NASA JPL RSA No.~1657376 and from NASA ADAP award 80NSSC19K1443. M.G. acknowledges partial support by NASA Chandra GO8-19104X/GO9-20114X and {\it HST} GO-15890.020/023-A, and the {\it BlackHoleWeather} program. E.N.E. would like to thank Bogazici University BAP for financial support under the project No 13760.

\bibliography{A3395}{}

\begin{thebibliography}{}
\expandafter\ifx\csname natexlab\endcsname\relax\def\natexlab#1{#1}\fi
\providecommand{\url}[1]{\href{#1}{#1}}
\providecommand{\dodoi}[1]{doi:~\href{http://doi.org/#1}{\nolinkurl{#1}}}
\providecommand{\doeprint}[1]{\href{http://ascl.net/#1}{\nolinkurl{http://ascl.net/#1}}}
\providecommand{\doarXiv}[1]{\href{https://arxiv.org/abs/#1}{\nolinkurl{https://arxiv.org/abs/#1}}}

\bibitem[{{Akamatsu} {et~al.}(2017){Akamatsu}, {Fujita}, {Akahori}, {Ishisaki},
  {Hayashida}, {Hoshino}, {Mernier}, {Yoshikawa}, {Sato}, \&
  {Kaastra}}]{akamatsu17}
{Akamatsu}, H., {Fujita}, Y., {Akahori}, T., {et~al.} 2017, \aap, 606, A1,
  \dodoi{10.1051/0004-6361/201730497}

\bibitem[{{Alvarez} {et~al.}(2018){Alvarez}, {Randall}, {Bourdin}, {Jones}, \&
  {Holley-Bockelmann}}]{alvarez18}
{Alvarez}, G.~E., {Randall}, S.~W., {Bourdin}, H., {Jones}, C., \&
  {Holley-Bockelmann}, K. 2018, \apj, 858, 44, \dodoi{10.3847/1538-4357/aabad0}

\bibitem[{{Anders} \& {Grevesse}(1989)}]{anders89}
{Anders}, E., \& {Grevesse}, N. 1989, \gca, 53, 197,
  \dodoi{10.1016/0016-7037(89)90286-X}

\bibitem[{{Arnaud}(1996)}]{arnaud96}
{Arnaud}, K.~A. 1996, in Astronomical Society of the Pacific Conference Series,
  Vol. 101, Astronomical Data Analysis Software and Systems V, ed. G.~H.
  {Jacoby} \& J.~{Barnes}, 17

\bibitem[{{Bond} {et~al.}(1996){Bond}, {Kofman}, \& {Pogosyan}}]{bond96}
{Bond}, J.~R., {Kofman}, L., \& {Pogosyan}, D. 1996, \nat, 380, 603,
  \dodoi{10.1038/380603a0}

\bibitem[{{Bonjean} {et~al.}(2018){Bonjean}, {Aghanim}, {Salom{\'e}},
  {Douspis}, \& {Beelen}}]{bonjean18}
{Bonjean}, V., {Aghanim}, N., {Salom{\'e}}, P., {Douspis}, M., \& {Beelen}, A.
  2018, \aap, 609, A49, \dodoi{10.1051/0004-6361/201731699}

\bibitem[{{Bourdin} {et~al.}(2020){Bourdin}, {Baldi}, {Kozmanyan}, \&
  {Mazzotta}}]{bourdin20}
{Bourdin}, H., {Baldi}, A.~S., {Kozmanyan}, A., \& {Mazzotta}, P. 2020, in
  European Physical Journal Web of Conferences, Vol. 228, European Physical
  Journal Web of Conferences, 00007, \dodoi{10.1051/epjconf/202022800007}

\bibitem[{{Cash}(1979)}]{cash79}
{Cash}, W. 1979, \apj, 228, 939, \dodoi{10.1086/156922}

\bibitem[{{Cautun} {et~al.}(2014){Cautun}, {van de Weygaert}, {Jones}, \&
  {Frenk}}]{cautun14}
{Cautun}, M., {van de Weygaert}, R., {Jones}, B. J.~T., \& {Frenk}, C.~S. 2014,
  \mnras, 441, 2923, \dodoi{10.1093/mnras/stu768}

\bibitem[{{Codis} {et~al.}(2012){Codis}, {Pichon}, {Devriendt}, {Slyz},
  {Pogosyan}, {Dubois}, \& {Sousbie}}]{codis12}
{Codis}, S., {Pichon}, C., {Devriendt}, J., {et~al.} 2012, \mnras, 427, 3320,
  \dodoi{10.1111/j.1365-2966.2012.21636.x}

\bibitem[{{De Grandi} {et~al.}(1999){De Grandi}, {Guzzo}, {B{\"o}hringer},
  {Molendi}, {Chincarini}, {Collins}, {Cruddace}, {Neumann}, {Schindler},
  {Schuecker}, \& {Voges}}]{degrandi99}
{De Grandi}, S., {Guzzo}, L., {B{\"o}hringer}, H., {et~al.} 1999, \apjl, 513,
  L17, \dodoi{10.1086/311900}

\bibitem[{{Donnelly} {et~al.}(2001){Donnelly}, {Forman}, {Jones}, {Quintana},
  {Ramirez}, {Churazov}, \& {Gilfanov}}]{donnely01}
{Donnelly}, R.~H., {Forman}, W., {Jones}, C., {et~al.} 2001, \apj, 562, 254,
  \dodoi{10.1086/323521}

\bibitem[{{Flin}(2003)}]{flin03}
{Flin}, P. 2003, Astronomical and Astrophysical Transactions, 22, 841,
  \dodoi{10.1080/1055679031000148677}

\bibitem[{{Gaspari}(2015)}]{gaspari15}
{Gaspari}, M. 2015, \mnras, 451, L60, \dodoi{10.1093/mnrasl/slv067}

\bibitem[{{Gaspari} {et~al.}(2014){Gaspari}, {Brighenti}, {Temi}, \&
  {Ettori}}]{gaspari14}
{Gaspari}, M., {Brighenti}, F., {Temi}, P., \& {Ettori}, S. 2014, \apjl, 783,
  L10, \dodoi{10.1088/2041-8205/783/1/L10}

\bibitem[{{Gaspari} \& {Churazov}(2013)}]{gaspari13}
{Gaspari}, M., \& {Churazov}, E. 2013, \aap, 559, A78,
  \dodoi{10.1051/0004-6361/201322295}

\bibitem[{{Gaspari} {et~al.}(2018){Gaspari}, {McDonald}, {Hamer},
  {et~al.}}]{gaspari18}
{Gaspari}, M., {McDonald}, M., {Hamer}, S.~L., {et~al.} 2018, \apj, 854, 167,
  \dodoi{10.3847/1538-4357/aaaa1b}

\bibitem[{{Gaspari} {et~al.}(2020){Gaspari}, {Tombesi}, \& {Cappi}}]{gaspari20}
{Gaspari}, M., {Tombesi}, F., \& {Cappi}, M. 2020, Nature Astronomy, 4, 10,
  \dodoi{10.1038/s41550-019-0970-1}

\bibitem[{{Gaspari} {et~al.}(2019){Gaspari}, {Eckert}, {Ettori}, {Tozzi},
  {Bassini}, {Rasia}, {Brighenti}, {Sun}, {Borgani}, {Johnson}, {Tremblay},
  {Stone}, {Temi}, {Yang}, {Tombesi}, \& {Cappi}}]{gaspari19}
{Gaspari}, M., {Eckert}, D., {Ettori}, S., {et~al.} 2019, \apj, 884, 169,
  \dodoi{10.3847/1538-4357/ab3c5d}

\bibitem[{{Gastaldello} {et~al.}(2015){Gastaldello}, {Wik}, {Molendi},
  {Westergaard}, {Hornstrup}, {Madejski}, {Ferreira}, {Boggs}, {Christensen},
  {Craig}, {Grefenstette}, {Hailey}, {Harrison}, {Madsen}, {Stern}, \&
  {Zhang}}]{gastaldello15}
{Gastaldello}, F., {Wik}, D.~R., {Molendi}, S., {et~al.} 2015, \apj, 800, 139,
  \dodoi{10.1088/0004-637X/800/2/139}

\bibitem[{{Gitti} {et~al.}(2010){Gitti}, {O'Sullivan}, {Giacintucci}, {David},
  {Vrtilek}, {Raychaudhury}, \& {Nulsen}}]{gitti10}
{Gitti}, M., {O'Sullivan}, E., {Giacintucci}, S., {et~al.} 2010, \apj, 714,
  758, \dodoi{10.1088/0004-637X/714/1/758}

\bibitem[{Hahn {et~al.}(2007)Hahn, Porciani, Carollo, \& Dekel}]{hahn07}
Hahn, O., Porciani, C., Carollo, C.~M., \& Dekel, A. 2007, Monthly Notices of
  the Royal Astronomical Society, 375, 489

\bibitem[{{Harrison} {et~al.}(2013){Harrison}, {Craig}, {Christensen},
  {Hailey}, {Zhang}, {Boggs}, {Stern}, {Cook}, {Forster}, {Giommi},
  {Grefenstette}, {Kim}, {Kitaguchi}, {Koglin}, {Madsen}, {Mao}, {Miyasaka},
  {Mori}, {Perri}, {Pivovaroff}, {Puccetti}, {Rana}, {Westergaard}, {Willis},
  {Zoglauer}, {An}, {Bachetti}, {Barri{\`e}re}, {Bellm}, {Bhalerao},
  {Brejnholt}, {Fuerst}, {Liebe}, {Markwardt}, {Nynka}, {Vogel}, {Walton},
  {Wik}, {Alexander}, {Cominsky}, {Hornschemeier}, {Hornstrup}, {Kaspi},
  {Madejski}, {Matt}, {Molendi}, {Smith}, {Tomsick}, {Ajello}, {Ballantyne},
  {Balokovi{\'c}}, {Barret}, {Bauer}, {Blandford}, {Brandt}, {Brenneman},
  {Chiang}, {Chakrabarty}, {Chenevez}, {Comastri}, {Dufour}, {Elvis}, {Fabian},
  {Farrah}, {Fryer}, {Gotthelf}, {Grindlay}, {Helfand}, {Krivonos}, {Meier},
  {Miller}, {Natalucci}, {Ogle}, {Ofek}, {Ptak}, {Reynolds}, {Rigby},
  {Tagliaferri}, {Thorsett}, {Treister}, \& {Urry}}]{harrison13}
{Harrison}, F.~A., {Craig}, W.~W., {Christensen}, F.~E., {et~al.} 2013, \apj,
  770, 103, \dodoi{10.1088/0004-637X/770/2/103}

\bibitem[{{Henry} \& {Briel}(1995)}]{henry95}
{Henry}, J.~P., \& {Briel}, U.~G. 1995, \apjl, 443, L9, \dodoi{10.1086/187823}

\bibitem[{{Hincks} {et~al.}(2022){Hincks}, {Radiconi}, {Romero},
  {Madhavacheril}, {Mroczkowski}, {Austermann}, {Barbavara}, {Battaglia},
  {Battistelli}, {Bond}, {Calabrese}, {de Bernardis}, {Devlin}, {Dicker},
  {Duff}, {Duivenvoorden}, {Dunkley}, {D{\"u}nner}, {Gallardo}, {Govoni},
  {Hill}, {Hilton}, {Hubmayr}, {Hughes}, {Lamagna}, {Lokken}, {Masi}, {Mason},
  {McMahon}, {Moodley}, {Murgia}, {Naess}, {Page}, {Piacentini}, {Salatino},
  {Sarazin}, {Schillaci}, {Sievers}, {Sif{\'o}n}, {Staggs}, {Ullom}, {Vacca},
  {Van Engelen}, {Vissers}, {Wollack}, \& {Xu}}]{hincks22}
{Hincks}, A.~D., {Radiconi}, F., {Romero}, C., {et~al.} 2022, \mnras, 510,
  3335, \dodoi{10.1093/mnras/stab3391}

\bibitem[{{Ishisaki} {et~al.}(2007){Ishisaki}, {Maeda}, {Fujimoto}, {Ozaki},
  {Ebisawa}, {Takahashi}, {Ueda}, {Ogasaka}, {Ptak}, {Mukai}, {Hamaguchi},
  {Hirayama}, {Kotani}, {Kubo}, {Shibata}, {Ebara}, {Furuzawa}, {Iizuka},
  {Inoue}, {Mori}, {Okada}, {Yokoyama}, {Matsumoto}, {Nakajima}, {Yamaguchi},
  {Anabuki}, {Tawa}, {Nagai}, {Katsuda}, {Hayashida}, {Bamba}, {Miller},
  {Sato}, \& {Yamasaki}}]{ishisaki07}
{Ishisaki}, Y., {Maeda}, Y., {Fujimoto}, R., {et~al.} 2007, \pasj, 59, 113,
  \dodoi{10.1093/pasj/59.sp1.S113}

\bibitem[{{Jones} {et~al.}(2009){Jones}, {Read}, {Saunders}, {Colless},
  {Jarrett}, {Parker}, {Fairall}, {Mauch}, {Sadler}, {Watson}, {Burton},
  {Campbell}, {Cass}, {Croom}, {Dawe}, {Fiegert}, {Frankcombe}, {Hartley},
  {Huchra}, {James}, {Kirby}, {Lahav}, {Lucey}, {Mamon}, {Moore}, {Peterson},
  {Prior}, {Proust}, {Russell}, {Safouris}, {Wakamatsu}, {Westra}, \&
  {Williams}}]{jones09}
{Jones}, D.~H., {Read}, M.~A., {Saunders}, W., {et~al.} 2009, \mnras, 399, 683,
  \dodoi{10.1111/j.1365-2966.2009.15338.x}

\bibitem[{{Kalberla} {et~al.}(2005){Kalberla}, {Burton}, {Hartmann}, {Arnal},
  {Bajaja}, {Morras}, \& {P{\"o}ppel}}]{kalberla05}
{Kalberla}, P.~M.~W., {Burton}, W.~B., {Hartmann}, D., {et~al.} 2005, \aap,
  440, 775, \dodoi{10.1051/0004-6361:20041864}

\bibitem[{{Koyama} {et~al.}(2007){Koyama}, {Tsunemi}, {Dotani}, {Bautz},
  {Hayashida}, {Tsuru}, {Matsumoto}, {Ogawara}, {Ricker}, {Doty}, {Kissel},
  {Foster}, {Nakajima}, {Yamaguchi}, {Mori}, {Sakano}, {Hamaguchi},
  {Nishiuchi}, {Miyata}, {Torii}, {Namiki}, {Katsuda}, {Matsuura}, {Miyauchi},
  {Anabuki}, {Tawa}, {Ozaki}, {Murakami}, {Maeda}, {Ichikawa}, {Prigozhin},
  {Boughan}, {Lamarr}, {Miller}, {Burke}, {Gregory}, {Pillsbury}, {Bamba},
  {Hiraga}, {Senda}, {Katayama}, {Kitamoto}, {Tsujimoto}, {Kohmura}, {Tsuboi},
  \& {Awaki}}]{koyama07}
{Koyama}, K., {Tsunemi}, H., {Dotani}, T., {et~al.} 2007, \pasj, 59, 23,
  \dodoi{10.1093/pasj/59.sp1.S23}

\bibitem[{{Kraljic} {et~al.}(2018){Kraljic}, {Arnouts}, {Pichon}, {Laigle}, {de
  la Torre}, {Vibert}, {Cadiou}, {Dubois}, {Treyer}, {Schimd}, {Codis}, {de
  Lapparent}, {Devriendt}, {Hwang}, {Le Borgne}, {Malavasi}, {Milliard},
  {Musso}, {Pogosyan}, {Alpaslan}, {Bland-Hawthorn}, \& {Wright}}]{kraljic18}
{Kraljic}, K., {Arnouts}, S., {Pichon}, C., {et~al.} 2018, \mnras, 474, 547,
  \dodoi{10.1093/mnras/stx2638}

\bibitem[{{Kuchner} {et~al.}(2020){Kuchner}, {Arag{\'o}n-Salamanca}, {Pearce},
  {Gray}, {Rost}, {Mu}, {Welker}, {Cui}, {Haggar}, {Laigle}, {Knebe},
  {Kraljic}, {Sarron}, \& {Yepes}}]{kuchner20}
{Kuchner}, U., {Arag{\'o}n-Salamanca}, A., {Pearce}, F.~R., {et~al.} 2020,
  \mnras, 494, 5473, \dodoi{10.1093/mnras/staa1083}

\bibitem[{{Kull} \& {B{\"o}hringer}(1999)}]{kull99}
{Kull}, A., \& {B{\"o}hringer}, H. 1999, \aap, 341, 23.
\newblock \doarXiv{astro-ph/9812319}

\bibitem[{{Laigle} {et~al.}(2015){Laigle}, {Pichon}, {Codis}, {Dubois}, {Le
  Borgne}, {Pogosyan}, {Devriendt}, {Peirani}, {Prunet}, {Rouberol}, {Slyz}, \&
  {Sousbie}}]{laigle15}
{Laigle}, C., {Pichon}, C., {Codis}, S., {et~al.} 2015, \mnras, 446, 2744,
  \dodoi{10.1093/mnras/stu2289}

\bibitem[{{Lakhchaura} {et~al.}(2011){Lakhchaura}, {Singh}, {Saikia}, \&
  {Hunstead}}]{lakhchaura11}
{Lakhchaura}, K., {Singh}, K.~P., {Saikia}, D.~J., \& {Hunstead}, R.~W. 2011,
  \apj, 743, 78, \dodoi{10.1088/0004-637X/743/1/78}

\bibitem[{{Lodders} {et~al.}(2009){Lodders}, {Palme}, \& {Gail}}]{lodders09}
{Lodders}, K., {Palme}, H., \& {Gail}, H.~P. 2009, LanB, 4B, 712,
  \dodoi{10.1007/978-3-540-88055-4\_34}

\bibitem[{{Lovisari} {et~al.}(2021){Lovisari}, {Ettori}, {Gaspari}, \&
  {Giles}}]{lovisari21}
{Lovisari}, L., {Ettori}, S., {Gaspari}, M., \& {Giles}, P.~A. 2021, Universe,
  7, 139, \dodoi{10.3390/universe7050139}

\bibitem[{{Madsen} {et~al.}(2017){Madsen}, {Christensen}, {Craig}, {Forster},
  {Grefenstette}, {Harrison}, {Miyasaka}, \& {Rana}}]{madsen17}
{Madsen}, K.~K., {Christensen}, F.~E., {Craig}, W.~W., {et~al.} 2017, Journal
  of Astronomical Telescopes, Instruments, and Systems, 3, 044003,
  \dodoi{10.1117/1.JATIS.3.4.044003}

\bibitem[{{Markevitch} {et~al.}(1998){Markevitch}, {Forman}, {Sarazin}, \&
  {Vikhlinin}}]{markevitch98}
{Markevitch}, M., {Forman}, W.~R., {Sarazin}, C.~L., \& {Vikhlinin}, A. 1998,
  \apj, 503, 77, \dodoi{10.1086/305976}

\bibitem[{{Markevitch} {et~al.}(1999){Markevitch}, {Sarazin}, \&
  {Vikhlinin}}]{markevitch99}
{Markevitch}, M., {Sarazin}, C.~L., \& {Vikhlinin}, A. 1999, \apj, 521, 526,
  \dodoi{10.1086/307598}

\bibitem[{{Markevitch} \& {Vikhlinin}(2007)}]{markevitch07}
{Markevitch}, M., \& {Vikhlinin}, A. 2007, \physrep, 443, 1,
  \dodoi{10.1016/j.physrep.2007.01.001}

\bibitem[{{Marziani} {et~al.}(2017){Marziani}, {D'Onofrio}, {Bettoni},
  {Poggianti}, {Moretti}, {Fasano}, {Fritz}, {Cava}, {Varela}, \&
  {Omizzolo}}]{marziani17}
{Marziani}, P., {D'Onofrio}, M., {Bettoni}, D., {et~al.} 2017, \aap, 599, A83,
  \dodoi{10.1051/0004-6361/201628941}

\bibitem[{{Mitsuda} {et~al.}(2007){Mitsuda}, {Bautz}, {Inoue}, {Kelley},
  {Koyama}, {Kunieda}, {Makishima}, {Ogawara}, {Petre}, {Takahashi}, {Tsunemi},
  {White}, {Anabuki}, {Angelini}, {Arnaud}, {Awaki}, {Bamba}, {Boyce}, {Brown},
  {Chan}, {Cottam}, {Dotani}, {Doty}, {Ebisawa}, {Ezoe}, {Fabian}, {Figueroa},
  {Fujimoto}, {Fukazawa}, {Furusho}, {Furuzawa}, {Gendreau}, {Griffiths},
  {Haba}, {Hamaguchi}, {Harrus}, {Hasinger}, {Hatsukade}, {Hayashida}, {Henry},
  {Hiraga}, {Holt}, {Hornschemeier}, {Hughes}, {Hwang}, {Ishida}, {Ishisaki},
  {Isobe}, {Itoh}, {Iyomoto}, {Kahn}, {Kamae}, {Katagiri}, {Kataoka},
  {Katayama}, {Kawai}, {Kilbourne}, {Kinugasa}, {Kissel}, {Kitamoto}, {Kohama},
  {Kohmura}, {Kokubun}, {Kotani}, {Kotoku}, {Kubota}, {Madejski}, {Maeda},
  {Makino}, {Markowitz}, {Matsumoto}, {Matsumoto}, {Matsuoka}, {Matsushita},
  {McCammon}, {Mihara}, {Misaki}, {Miyata}, {Mizuno}, {Mori}, {Mori}, {Morii},
  {Moseley}, {Mukai}, {Murakami}, {Murakami}, {Mushotzky}, {Nagase}, {Namiki},
  {Negoro}, {Nakazawa}, {Nousek}, {Okajima}, {Ogasaka}, {Ohashi}, {Oshima},
  {Ota}, {Ozaki}, {Ozawa}, {Parmar}, {Pence}, {Porter}, {Reeves}, {Ricker},
  {Sakurai}, {Sanders}, {Senda}, {Serlemitsos}, {Shibata}, {Soong}, {Smith},
  {Suzuki}, {Szymkowiak}, {Takahashi}, {Tamagawa}, {Tamura}, {Tamura},
  {Tanaka}, {Tashiro}, {Tawara}, {Terada}, {Terashima}, {Tomida}, {Torii},
  {Tsuboi}, {Tsujimoto}, {Tsuru}, {Turner}, {Ueda}, {Ueno}, {Ueno}, {Uno},
  {Urata}, {Watanabe}, {Yamamoto}, {Yamaoka}, {Yamasaki}, {Yamashita},
  {Yamauchi}, {Yamauchi}, {Yaqoob}, {Yonetoku}, \& {Yoshida}}]{mitsuda07}
{Mitsuda}, K., {Bautz}, M., {Inoue}, H., {et~al.} 2007, \pasj, 59, 1,
  \dodoi{10.1093/pasj/59.sp1.S1}

\bibitem[{{Munro} \& {Dubois}(1995)}]{munro95}
{Munro}, D.~H., \& {Dubois}, P.~F. 1995, Computers in Physics, 9, 609,
  \dodoi{10.1063/1.4823451}

\bibitem[{{Nakamura} {et~al.}(1995){Nakamura}, {Hattori}, \&
  {Mineshige}}]{nakamura95}
{Nakamura}, F.~E., {Hattori}, M., \& {Mineshige}, S. 1995, \aap, 302, 649.
\newblock \doarXiv{astro-ph/9505004}

\bibitem[{{Paturel} {et~al.}(2003){Paturel}, {Petit}, {Prugniel}, {Theureau},
  {Rousseau}, {Brouty}, {Dubois}, \& {Cambr{\'e}sy}}]{paturel03}
{Paturel}, G., {Petit}, C., {Prugniel}, P., {et~al.} 2003, \aap, 412, 45,
  \dodoi{10.1051/0004-6361:20031411}

\bibitem[{{Piffaretti} {et~al.}(2005){Piffaretti}, {Jetzer}, {Kaastra}, \&
  {Tamura}}]{piffaretti05}
{Piffaretti}, R., {Jetzer}, P., {Kaastra}, J.~S., \& {Tamura}, T. 2005, \aap,
  433, 101, \dodoi{10.1051/0004-6361:20041888}

\bibitem[{{Planck Collaboration} {et~al.}(2013){Planck Collaboration}, {Ade},
  {Aghanim}, {Arnaud}, {Ashdown}, {Atrio-Barandela}, {Aumont}, {Baccigalupi},
  {Balbi}, {Banday}, {Barreiro}, {Battaner}, {Benabed}, {Beno{\^\i}t},
  {Bernard}, {Bersanelli}, {Bhatia}, {Bikmaev}, {B{\"o}hringer}, {Bonaldi},
  {Bond}, {Borrill}, {Bouchet}, {Bourdin}, {Burenin}, {Burigana}, {Cabella},
  {Cardoso}, {Castex}, {Catalano}, {Cay{\'o}n}, {Chamballu}, {Chary}, {Chiang},
  {Chon}, {Christensen}, {Clements}, {Colafrancesco}, {Colombo}, {Comis},
  {Coulais}, {Crill}, {Cuttaia}, {Danese}, {Davis}, {de Bernardis}, {de
  Gasperis}, {de Zotti}, {Delabrouille}, {D{\'e}mocl{\`e}s}, {D{\'e}sert},
  {Diego}, {Dolag}, {Dole}, {Donzelli}, {Dor{\'e}}, {D{\"o}rl}, {Douspis},
  {Dupac}, {Efstathiou}, {En{\ss}lin}, {Eriksen}, {Finelli}, {Flores-Cacho},
  {Forni}, {Frailis}, {Franceschi}, {Frommert}, {Ganga}, {G{\'e}nova-Santos},
  {Giard}, {Gilfanov}, {Giraud-H{\'e}raud}, {Gonz{\'a}lez-Nuevo}, {G{\'o}rski},
  {Gregorio}, {Gruppuso}, {Hansen}, {Harrison}, {Hempel},
  {Henrot-Versill{\'e}}, {Hern{\'a}ndez-Monteagudo}, {Herranz}, {Hildebrandt},
  {Hivon}, {Hobson}, {Holmes}, {Hovest}, {Hurier}, {Jaffe}, {Jaffe},
  {Jagemann}, {Jones}, {Juvela}, {Khamitov}, {Kisner}, {Kneissl}, {Knoche},
  {Knox}, {Kunz}, {Kurki-Suonio}, {Lagache}, {Lamarre}, {Lasenby}, {Lawrence},
  {Le Jeune}, {Leonardi}, {Lilje}, {Linden-V{\o}rnle}, {L{\'o}pez-Caniego},
  {Lubin}, {Luzzi}, {Mac{\'\i}as-P{\'e}rez}, {Maffei}, {Maino}, {Mandolesi},
  {Maris}, {Marleau}, {Marshall}, {Mart{\'\i}nez-Gonz{\'a}lez}, {Masi},
  {Massardi}, {Matarrese}, {Matthai}, {Mazzotta}, {Mei}, {Melchiorri}, {Melin},
  {Mendes}, {Mennella}, {Mitra}, {Miville-Desch{\`e}nes}, {Moneti}, {Montier},
  {Morgante}, {Munshi}, {Murphy}, {Naselsky}, {Nati}, {Natoli},
  {N{\o}rgaard-Nielsen}, {Noviello}, {Novikov}, {Novikov}, {Osborne}, {Pajot},
  {Paoletti}, {Pasian}, {Patanchon}, {Perdereau}, {Perotto}, {Perrotta},
  {Piacentini}, {Piat}, {Pierpaoli}, {Piffaretti}, {Plaszczynski},
  {Pointecouteau}, {Polenta}, {Ponthieu}, {Popa}, {Poutanen}, {Pratt},
  {Prunet}, {Puget}, {Rachen}, {Rebolo}, {Reinecke}, {Remazeilles}, {Renault},
  {Ricciardi}, {Riller}, {Ristorcelli}, {Rocha}, {Roman}, {Rosset}, {Rossetti},
  {Rubi{\~n}o-Mart{\'\i}n}, {Rusholme}, {Sandri}, {Savini}, {Schaefer},
  {Scott}, {Smoot}, {Starck}, {Sudiwala}, {Sunyaev}, {Sutton}, {Suur-Uski},
  {Sygnet}, {Tauber}, {Terenzi}, {Toffolatti}, {Tomasi}, {Tristram}, {Tucci},
  {Valenziano}, {Van Tent}, {Vielva}, {Villa}, {Vittorio}, {Wade}, {Wandelt},
  {Welikala}, {White}, {Yvon}, {Zacchei}, \& {Zonca}}]{planck13}
{Planck Collaboration}, {Ade}, P.~A.~R., {Aghanim}, N., {et~al.} 2013, \aap,
  550, A134, \dodoi{10.1051/0004-6361/201220194}

\bibitem[{{Pratt} {et~al.}(2006){Pratt}, {Arnaud}, \&
  {Pointecouteau}}]{pratt06}
{Pratt}, G.~W., {Arnaud}, M., \& {Pointecouteau}, E. 2006, \aap, 446, 429,
  \dodoi{10.1051/0004-6361:20054025}

\bibitem[{{Reiprich} {et~al.}(2021){Reiprich}, {Veronica}, {Pacaud},
  {Ramos-Ceja}, {Ota}, {Sanders}, {Kara}, {Erben}, {Klein}, {Erler}, {Kerp},
  {Hoang}, {Br{\"u}ggen}, {Marvil}, {Rudnick}, {Biffi}, {Dolag},
  {Aschersleben}, {Basu}, {Brunner}, {Bulbul}, {Dennerl}, {Eckert}, {Freyberg},
  {Gatuzz}, {Ghirardini}, {K{\"a}fer}, {Merloni}, {Migkas}, {Nandra},
  {Predehl}, {Robrade}, {Salvato}, {Whelan}, {Diaz-Ocampo}, {Hernandez-Lang},
  {Zenteno}, {Brown}, {Collier}, {Diego}, {Hopkins}, {Kapinska}, {Koribalski},
  {Mroczkowski}, {Norris}, {O'Brien}, \& {Vardoulaki}}]{erosita21}
{Reiprich}, T.~H., {Veronica}, A., {Pacaud}, F., {et~al.} 2021, \aap, 647, A2,
  \dodoi{10.1051/0004-6361/202039590}

\bibitem[{{Rossetti} {et~al.}(2007){Rossetti}, {Ghizzardi}, {Molendi}, \&
  {Finoguenov}}]{rossetti07}
{Rossetti}, M., {Ghizzardi}, S., {Molendi}, S., \& {Finoguenov}, A. 2007, \aap,
  463, 839, \dodoi{10.1051/0004-6361:20054621}

\bibitem[{{Sarazin}(1986)}]{sarazin86}
{Sarazin}, C.~L. 1986, Reviews of Modern Physics, 58, 1,
  \dodoi{10.1103/RevModPhys.58.1}

\bibitem[{{Schellenberger} {et~al.}(2015){Schellenberger}, {Reiprich},
  {Lovisari}, {Nevalainen}, \& {David}}]{schellenberger15}
{Schellenberger}, G., {Reiprich}, T.~H., {Lovisari}, L., {Nevalainen}, J., \&
  {David}, L. 2015, \aap, 575, A30, \dodoi{10.1051/0004-6361/201424085}

\bibitem[{{Smith} {et~al.}(2001){Smith}, {Brickhouse}, {Liedahl}, \&
  {Raymond}}]{smith01}
{Smith}, R.~K., {Brickhouse}, N.~S., {Liedahl}, D.~A., \& {Raymond}, J.~C.
  2001, \apjl, 556, L91, \dodoi{10.1086/322992}

\bibitem[{{Somboonpanyakul} {et~al.}(2021){Somboonpanyakul}, {McDonald},
  {Bayliss}, {Voit}, {Donahue}, {Gaspari}, {Dahle}, {Rivera-Thorsen}, \&
  {Stark}}]{somboonpanyakul21}
{Somboonpanyakul}, T., {McDonald}, M., {Bayliss}, M., {et~al.} 2021, \apjl,
  907, L12, \dodoi{10.3847/2041-8213/abd540}

\bibitem[{{Str{\"u}der} {et~al.}(2001){Str{\"u}der}, {Briel}, {Dennerl},
  {Hartmann}, {Kendziorra}, {Meidinger}, {Pfeffermann}, {Reppin}, {Aschenbach},
  {Bornemann}, {Br{\"a}uninger}, {Burkert}, {Elender}, {Freyberg}, {Haberl},
  {Hartner}, {Heuschmann}, {Hippmann}, {Kastelic}, {Kemmer}, {Kettenring},
  {Kink}, {Krause}, {M{\"u}ller}, {Oppitz}, {Pietsch}, {Popp}, {Predehl},
  {Read}, {Stephan}, {St{\"o}tter}, {Tr{\"u}mper}, {Holl}, {Kemmer}, {Soltau},
  {St{\"o}tter}, {Weber}, {Weichert}, {von Zanthier}, {Carathanassis}, {Lutz},
  {Richter}, {Solc}, {B{\"o}ttcher}, {Kuster}, {Staubert}, {Abbey}, {Holland},
  {Turner}, {Balasini}, {Bignami}, {La Palombara}, {Villa}, {Buttler},
  {Gianini}, {Lain{\'e}}, {Lumb}, \& {Dhez}}]{struder01}
{Str{\"u}der}, L., {Briel}, U., {Dennerl}, K., {et~al.} 2001, \aap, 365, L18,
  \dodoi{10.1051/0004-6361:20000066}

\bibitem[{{Sugawara} {et~al.}(2017){Sugawara}, {Takizawa}, {Itahana},
  {Akamatsu}, {Fujita}, {Ohashi}, \& {Ishisaki}}]{sugawara17}
{Sugawara}, Y., {Takizawa}, M., {Itahana}, M., {et~al.} 2017, \pasj, 69, 93,
  \dodoi{10.1093/pasj/psx104}

\bibitem[{{Sun} {et~al.}(2007){Sun}, {Jones}, {Forman}, {Vikhlinin}, {Donahue},
  \& {Voit}}]{sun07}
{Sun}, M., {Jones}, C., {Forman}, W., {et~al.} 2007, \apj, 657, 197,
  \dodoi{10.1086/510895}

\bibitem[{{Tawa} {et~al.}(2008){Tawa}, {Hayashida}, {Nagai}, {Nakamoto},
  {Tsunemi}, {Yamaguchi}, {Ishisaki}, {Miller}, {Mizuno}, {Dotani}, {Ozaki}, \&
  {Katayama}}]{tawa08}
{Tawa}, N., {Hayashida}, K., {Nagai}, M., {et~al.} 2008, \pasj, 60, 11.
\newblock \doarXiv{0803.0616}

\bibitem[{{Tittley} \& {Henriksen}(2001)}]{tittley01}
{Tittley}, E.~R., \& {Henriksen}, M. 2001, \apj, 563, 673,
  \dodoi{10.1086/323955}

\bibitem[{{Tozzi} {et~al.}(2006){Tozzi}, {Gilli}, {Mainieri}, {Norman},
  {Risaliti}, {Rosati}, {Bergeron}, {Borgani}, {Giacconi}, {Hasinger},
  {Nonino}, {Streblyanska}, {Szokoly}, {Wang}, \& {Zheng}}]{tozzi06}
{Tozzi}, P., {Gilli}, R., {Mainieri}, V., {et~al.} 2006, \aap, 451, 457,
  \dodoi{10.1051/0004-6361:20042592}

\bibitem[{{T{\"u}mer} {et~al.}(2019){T{\"u}mer}, {Tombesi}, {Bourdin}, {Ercan},
  {Gaspari}, \& {Serafinelli}}]{tumer19}
{T{\"u}mer}, A., {Tombesi}, F., {Bourdin}, H., {et~al.} 2019, \aap, 629, A82,
  \dodoi{10.1051/0004-6361/201935660}

\bibitem[{{Turner} {et~al.}(2001){Turner}, {Abbey}, {Arnaud}, {Balasini},
  {Barbera}, {Belsole}, {Bennie}, {Bernard}, {Bignami}, {Boer}, {Briel},
  {Butler}, {Cara}, {Chabaud}, {Cole}, {Collura}, {Conte}, {Cros}, {Denby},
  {Dhez}, {Di Coco}, {Dowson}, {Ferrando}, {Ghizzardi}, {Gianotti}, {Goodall},
  {Gretton}, {Griffiths}, {Hainaut}, {Hochedez}, {Holland}, {Jourdain},
  {Kendziorra}, {Lagostina}, {Laine}, {La Palombara}, {Lortholary}, {Lumb},
  {Marty}, {Molendi}, {Pigot}, {Poindron}, {Pounds}, {Reeves}, {Reppin},
  {Rothenflug}, {Salvetat}, {Sauvageot}, {Schmitt}, {Sembay}, {Short},
  {Spragg}, {Stephen}, {Str{\"u}der}, {Tiengo}, {Trifoglio}, {Tr{\"u}mper},
  {Vercellone}, {Vigroux}, {Villa}, {Ward}, {Whitehead}, \& {Zonca}}]{turner01}
{Turner}, M.~J.~L., {Abbey}, A., {Arnaud}, M., {et~al.} 2001, \aap, 365, L27,
  \dodoi{10.1051/0004-6361:20000087}

\bibitem[{{Valageas} {et~al.}(2003){Valageas}, {Schaeffer}, \&
  {Silk}}]{valageas03}
{Valageas}, P., {Schaeffer}, R., \& {Silk}, J. 2003, \mnras, 344, 53,
  \dodoi{10.1046/j.1365-8711.2003.06781.x}

\bibitem[{{Voit} {et~al.}(2005){Voit}, {Kay}, \& {Bryan}}]{voit05}
{Voit}, G.~M., {Kay}, S.~T., \& {Bryan}, G.~L. 2005, \mnras, 364, 909,
  \dodoi{10.1111/j.1365-2966.2005.09621.x}

\bibitem[{{Werner} {et~al.}(2008){Werner}, {Finoguenov}, {Kaastra},
  {Simionescu}, {Dietrich}, {Vink}, \& {B{\"o}hringer}}]{werner08}
{Werner}, N., {Finoguenov}, A., {Kaastra}, J.~S., {et~al.} 2008, \aap, 482,
  L29, \dodoi{10.1051/0004-6361:200809599}

\bibitem[{{Westergaard}(2011)}]{westergaard11}
{Westergaard}, N.~J. 2011, in Society of Photo-Optical Instrumentation
  Engineers (SPIE) Conference Series, Vol. 8147, Society of Photo-Optical
  Instrumentation Engineers (SPIE) Conference Series, ed. S.~L. {O'Dell} \&
  G.~{Pareschi}, 81470Y, \dodoi{10.1117/12.895310}

\bibitem[{{Wik} {et~al.}(2014){Wik}, {Hornstrup}, {Molendi}, {Madejski},
  {Harrison}, {Zoglauer}, {Grefenstette}, {Gastaldello}, {Madsen},
  {Westergaard}, {Ferreira}, {Kitaguchi}, {Pedersen}, {Boggs}, {Christensen},
  {Craig}, {Hailey}, {Stern}, \& {Zhang}}]{wik14}
{Wik}, D.~R., {Hornstrup}, A., {Molendi}, S., {et~al.} 2014, \apj, 792, 48,
  \dodoi{10.1088/0004-637X/792/1/48}

\bibitem[{{Wilms} {et~al.}(2000){Wilms}, {Allen}, \& {McCray}}]{wilms00}
{Wilms}, J., {Allen}, A., \& {McCray}, R. 2000, \apj, 542, 914,
  \dodoi{10.1086/317016}

\bibitem[{{Wittor} \& {Gaspari}(2020)}]{wittor20}
{Wittor}, D., \& {Gaspari}, M. 2020, \mnras, 498, 4983,
  \dodoi{10.1093/mnras/staa2747}

\bibitem[{{Zinger} {et~al.}(2016){Zinger}, {Dekel}, {Birnboim}, {Kravtsov}, \&
  {Nagai}}]{zinger16}
{Zinger}, E., {Dekel}, A., {Birnboim}, Y., {Kravtsov}, A., \& {Nagai}, D. 2016,
  \mnras, 461, 412, \dodoi{10.1093/mnras/stw1283}

\end{thebibliography}
\bibliographystyle{aasjournal}

\appendix

\section{\nustar\ Background Spectra}\label{sec:bgdfit}
\begin{figure*}[h]
\centering
\includegraphics[width=83mm]{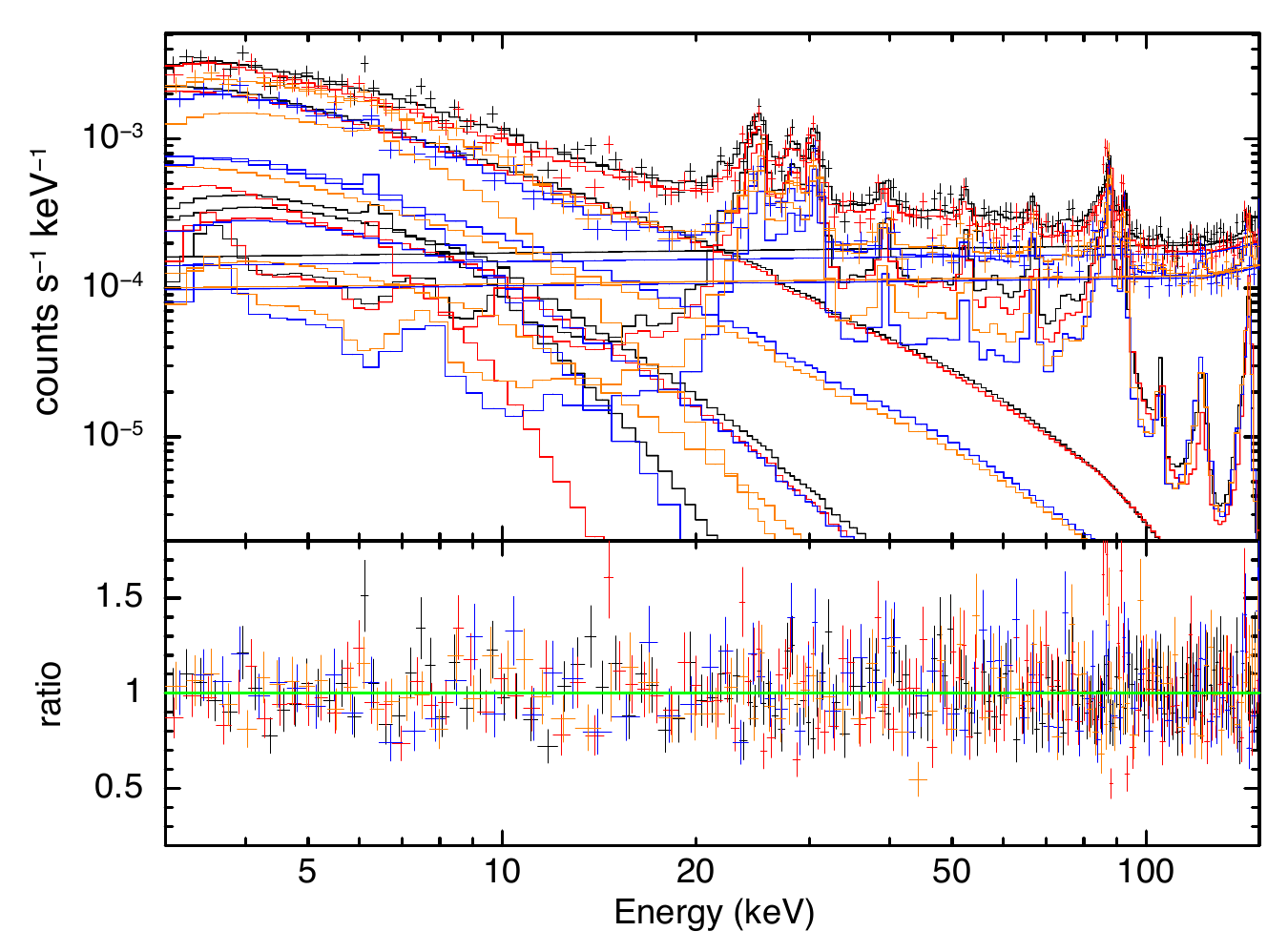}
\includegraphics[width=83mm]{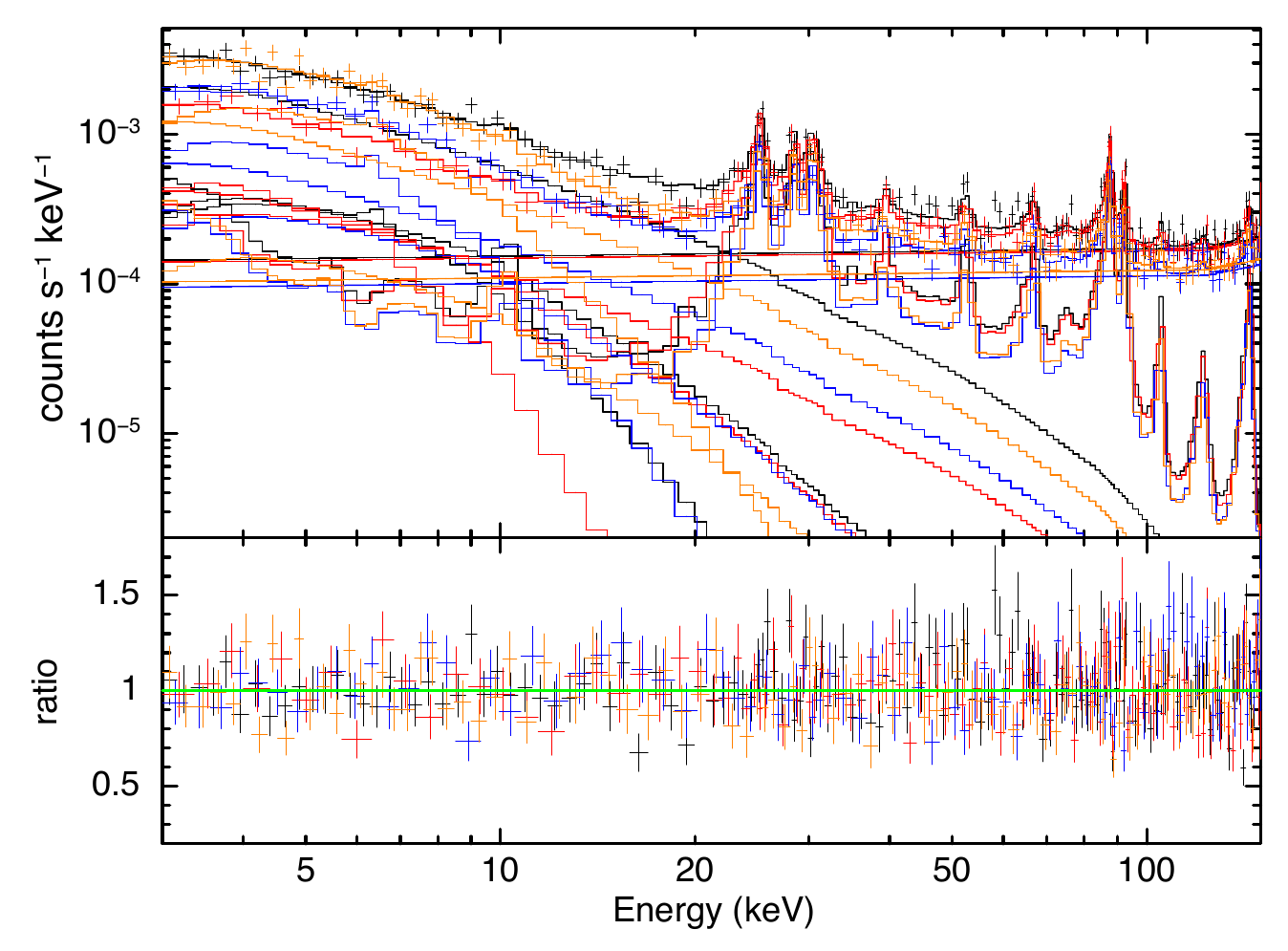}
\caption{Joint-fit of background and cluster emission of \nustar\ FPMA (\textit{left panel}) and FPMB (\textit{right panel}). Each color represents a region selected for the background fit. \label{fig:nuskybgd}}
\end{figure*}

In this appendix, we present the \nustar\ background spectral fit in Fig.~\ref{fig:nuskybgd}. The background assessment is described in detail in Section~\ref{sec:nustarbgd} and in \citet{wik14}.

\section{Results of the Ray-trace Simulation Spectral Fits results}\label{sec:raytracefit}

Here we present the spectral fits of the ray-traced single and double bounce photons extracted from regions of interest shown in Fig.~\ref{fig:6reg}. The fitting procedure is described in detail in Section~\ref{sec:SL}.

\begin{figure*}[ht!]
\centering
\includegraphics[width=180mm]{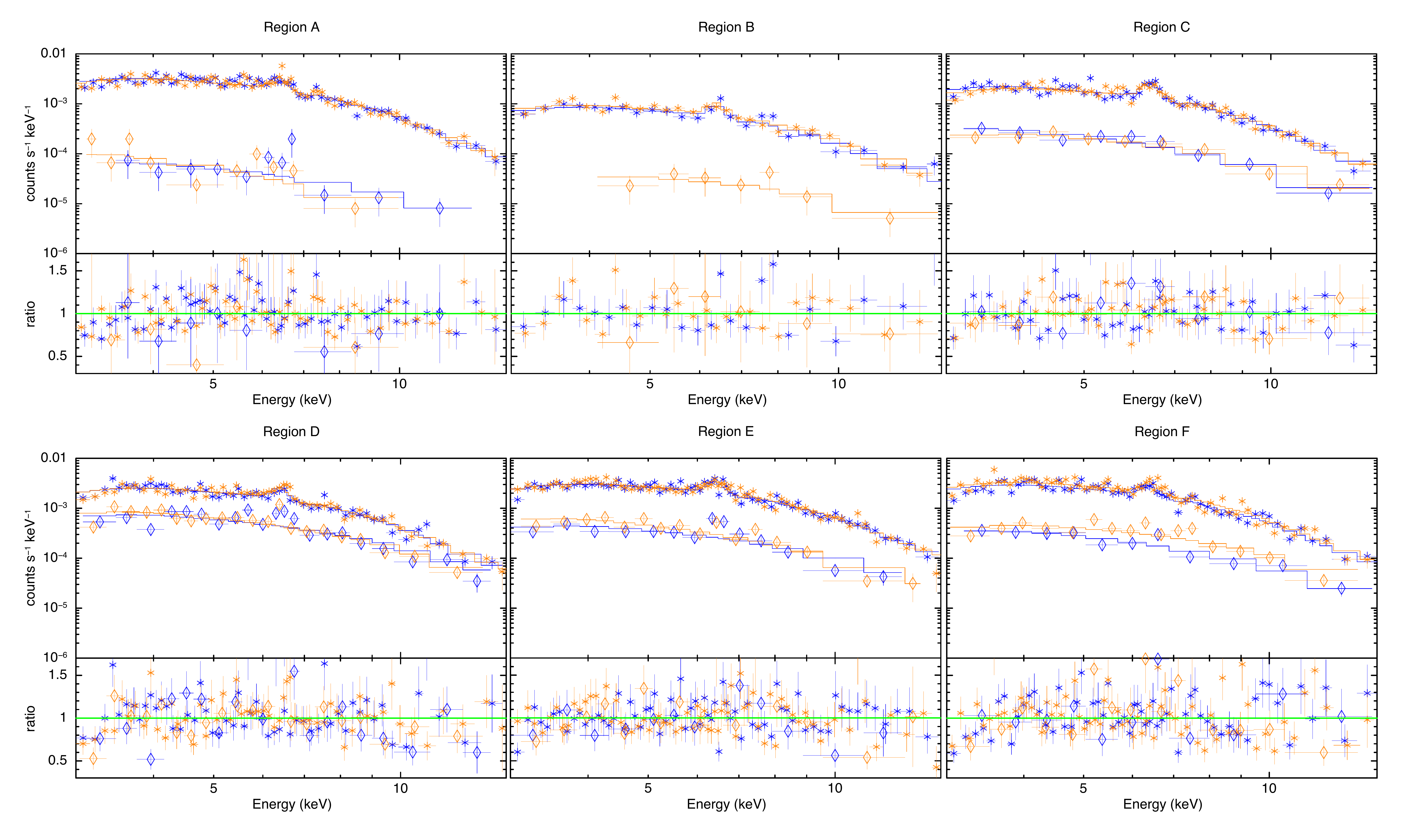}
\caption{Ray-trace simulation fits of spectra from the regions of interest shown in Fig.~\ref{fig:6reg} in the 3-15 keV band, where blue denotes FPMA, and orange denotes FPMB. Simulated single reflected photons are presented with diamond marker, and simulated double reflected photons with asterisk marker. For plotting purposes, adjacent bins are grouped until they have a significant detection at least as large as 5$\sigma$, with maximum 5 bins, except for the single reflected photons in regions A and B due to low counts. There were no data points from scattered light for FPMA for Region B between 3-15 keV, therefore is not shown. \label{fig:scatterfit}}
\end{figure*}

\section{Results of the Hot spot spectral Fits}\label{sec:hotspotfit}

The results of the spectral fit parameters from the circular region with r = 1.5$\arcmin$, centered centered at X shown in Fig.~\ref{fig:6reg}, are presented in this section. Here we have two tables, one for the \xmm\ data, and one for the joint \nustar\ and \xmm\ spectral analysis. 

We realized different spectral fit processes by both freeing and fixing abundance and {\it N$_{H}$} values to give the fitting procedure more independence to find a high temperature component. This analysis is described in detail in Section~\ref{sec:allfit}.

\begin{table*}[ht!]
\caption{Spectral parameters of \xmm\ analysis for the r = 1.5$\arcmin$ region centered at X shown in Fig.~\ref{fig:6reg} in 0.5-9.0 keV energy band. \texttt{apec} normalization (\textit{norm}) is given in $\frac{10^{-14}}{4\pi \left [ D_A(1+z) \right ]^2}\int n_{e}n_{H}dV$.}
\label{tab:centerXfitxmm}    
\centering
\begin{tabular}{l c c c c}
\hline\hline
 & \multicolumn{2}{c}{Fixed \textit{N$_{H}$}} & \multicolumn{2}{c}{Free \textit{N$_{H}$}}  \\
 \hline
  & Fixed \textit{Z} & Free \textit{Z} & Fixed \textit{Z} & Free \textit{Z}  \\
\hline
\textit{N$_{H}$} (10$^{20}$ cm$^{-2}$) & 6.3 & 6.3 & 8.51$^{+2.92}_{-4.59}$ & 11.0$^{+20.1}_{-6.11}$ \\  
\textit{kT} (keV)  & 4.79$^{+1.88}_{-1.10}$ & 5.13$^{+1.80}_{-1.48}$ & 4.43$^{+2.15}_{-1.63}$ & 3.98$^{+2.44}_{-1.16}$ \\  
\textit{Z} (\textit{Z$_{\odot}$})  & 0.23 & 0.03$^{+0.44}_{-0.03}$ & 0.23 & 0.05$^{+0.35}_{-0.05}$  \\  
\textit{norm} (10$^{-5}$ cm$^{-5}$)   & 2.95$^{+0.19}_{-0.21}$ & 3.10$^{+0.38}_{-0.23}$ & 3.06$^{+1.11}_{-0.36}$ & 3.43$^{+0.91}_{-0.60}$  \\  
\textit{$\chi{^2}$ / $\nu$}  & 181.39 / 97 & 180.27 / 96 & 180.52 / 96 & 179.13 / 95  \\  
\hline
\end{tabular}
\end{table*}

\begin{table*}[ht!]
\caption{Spectral parameters of \nustar\ (3.0-15.0 keV) and \xmm\ (0.5-9.0 keV) analysis for the r = 1.5$\arcmin$ region centered at X shown in Fig.~\ref{fig:6reg}. \texttt{apec} normalization (\textit{norm}) is given in $\frac{10^{-14}}{4\pi \left [ D_A(1+z) \right ]^2}\int n_{e}n_{H}dV$.}
\label{tab:centerXfitxmmnustar}    
\centering
\begin{tabular}{l c c c c}
\hline\hline
 & \multicolumn{2}{c}{Fixed \textit{N$_{H}$}} & \multicolumn{2}{c}{Free \textit{N$_{H}$}} \\
   \hline
 & Fixed \textit{Z} & Free \textit{Z} & Fixed \textit{Z} & Free \textit{Z}  \\
\hline
\textit{N$_{H}$} (10$^{20}$ cm$^{-2}$) & 6.3 & 6.3 & 14.8$^{+18.1}_{-6.31}$ & 13.7$^{+7.71}_{-5.81}$ \\  
\textit{kT} (keV)  & 3.75$^{+0.72}_{-0.46}$ & 3.97$^{+0.55}_{-0.53}$ & 3.58$^{+0.55}_{-0.56}$ & 3.74$^{+0.56}_{-0.51}$ \\  
\textit{Z} (\textit{Z$_{\odot}$}) & 0.23 & 0.09 (upper limit) & 0.23 & 0.10 (upper limit) \\  
\textit{norm} (10$^{-5}$ cm$^{-5}$) & 2.83$^{+0.16}_{-0.29}$ & 3.04$^{+0.19}_{-0.26}$ & 3.18$^{+0.65}_{-0.36}$ & 3.43$^{+0.54}_{-0.42}$ \\  
\textit{C / $\nu$} &  1334.21 / 1200 & 1321.82 / 1199 & 1322.23 / 1199 & 1316.23 / 1198 \\  
\hline
\\[-0.95em]
\end{tabular}
\end{table*}

\end{document}